\documentstyle[12pt]{article}
\begin{document}

\rightline{CWRU-P13-95}

\

\vskip 2 truecm

\begin{center}
{\large{\bf {
Aspects of ${}^3$He and the Standard Electroweak Model
}}}

\vskip 0.5 truecm

\obeylines{G.E. Volovik\footnote{volovik@boojum.hut.fi}
Low Temperature Laboratory,
Helsinki University of Technology
Otakaari 3A, 02150 Espoo, Finland
and
L.D. Landau Institute for Theoretical Physics,
Kosygin Str. 2, 117940 Moscow, Russia

\vskip 0.5 truecm

Tanmay Vachaspati\footnote{txv7@po.cwru.edu}
Physics Department, Case Western Reserve University
Cleveland, OH 44106, USA.
}
\end{center}

\vskip 1 truecm

\begin{abstract}
We describe certain aspects of ${}^3 He$ and compare them to related
aspects of the standard electroweak model of particle physics.
We note various
similarities in the order parameter structure, defect structure,
interactions with fermions and anomalies in the two systems. Many issues
in the condensed matter literature that are often confusing to the
particle physics reader and vice versa are clarified.
\end{abstract}

\vfill \eject

\tableofcontents

\section{Introduction}\label{intro}

It is well recognized that there are many similarities between
condensed matter systems and particle physics stemming from the
fact that both systems are described by field theory. Most of
the common phenomena - an example of which is spontaneous symmetry
breaking - are yet to be confirmed in the particle physics scenario
while they are quite routinely observed by condensed matter
physicists. Several phenomena occuring in condensed matter systems
are expected to also occur in speculative particle physics schemes.
These include the possibility of topological defects with their
potential astrophysical and cosmological consequences.

Particle physics as we know it today is accurately described by
the standard model. The model has been remarkably successful and
would almost universally be accepted if (and when) the Higgs particle
is found. At the same time, it should be noted that only the perturbative
features of the model have been put to the test and it would be
worthwhile to have a better understanding of the non-perturbative
aspects of the theory. Such an understanding may be crucial, for
example, for testing the hypothesis that the baryon number of the
universe was produced (baryogenesis) during the electroweak phase
transition in the early universe. The possibility of directly testing
the non-perturbative aspects of the standard model seem remote at
this time. In addition, some insight as to what to expect can help
in such attempts when the time comes. To gain some intuition, we
need to find a condensed matter system with some similarities to the
standard electroweak model.

$^3$He is a well-studied condensed matter system and has an
order parameter that has rich structure and yields
several different phases. The A-phase is particularly complex
and the transition of $^3$He to the A-phase closely resembles
the electroweak symmetry breaking. This
and other similarities between the standard electroweak model and
$^3$He have been noticed by one of us in earlier work \cite{Exotic}.
Here we shall extend this earlier work with clarifying
remarks and also point out differences in the two systems.
In particular, we discuss the vortices of $^3$He which are similar
to the $Z-$strings of the electroweak model \cite{nambu, tv}
and connect the baryon number
contained in configurations of $Z-$strings \cite{tvgf, jgtv}
with the mass current anomaly in $^3$He-A.

The superfluid $^3$He is a unique system among the other
condensed matter, because it has the maximum
symmetry breaking, which  can be compared only  with
the vacuum in the elementary particle physics. This results in a variety
of topological defects in $^3$He, such as monopoles, hedgehogs,
boojums, solitons, domain walls, textures,  quantized vortices, half-quantum
vortices (the counterpart  of the   Alice strings),   strings terminating on
monopoles and domain walls terminating on strings, etc., and in related
phenomena such as  topological confinement, symmetry breaking with
parity violation in the vortex core,  Aharonov-Bohm effect on global strings,
topological transitions, transitions mediated by monopoles and hedgehogs,
bosonic  and fermionic zero modes on vortices, genesis of the fermionic
charges by moving vortices,  vacuum polarization and
vacuum instability. This makes the
superfluid $^3$He a working laboratory for modelling different processes
which can occur in the physical vacuum.

\section{$^3$He and the electroweak model: similarities}

There are two ``levels'' on which $^3$He-A and the standard electroweak
model resemble each other:

{\bf Level 1:} The symmetry group of $^3$He-A and of the standard
electroweak model are very similar.

{\bf Level 2:} The interactions of the low energy fermions with the
$^3$He-A order parameter closely resembles the interactions of the fermions
in the electroweak model with the gauge fields present in the model.

\subsection{Broken symmetry} \label{Broken symmetry}

The symmetry group $G_{ew}$ of the standard electroweak
model is:
$$
G_{\rm ew}= [ SU(2)_{T}\times  U(1)_Y ]/Z_2 \ .
\eqno(2.1. 1a)$$
Here $SU(2)_T$ is the group of isotopic rotations with the generator ${\bf T}$.
The symmetry group $G_{\rm 3He}$ in liquid $^3$He
$$
G_{\rm 3He}=SO(3)_{S}\times SO(3)_{L}\times  U(1)_N \ ,
\eqno(2.1.1b)
$$
contains the groups of the independent
{\it orbital} and {\it spin}  rotations: the operator
${\bf L}$ is the generator of  the orbital rotations  $SO(3)_{L}$,
while the spin operator   ${\bf S}$   is the generator of the spin rotations
$SO(3)_{S}$. The operator ${\bf N}$ of $^3$He particle
number plays the part of hypercharge ${\bf Y}$.

After the electroweak transition the little symmetry group $H$ of the
electroweak vacuum is
$$
H_{\rm ew}=U(1)_Q\equiv U(1)_{T_3+Y/2} \ ,
\eqno(2.1.2a)
$$
where  ${\bf Q}={\bf T}_3+(1/2){\bf Y}$ is the generator
of electric charge.

After the superfluid transtition of $^3$He into the A-phase state the vacuum
becomes anisotropic. The spin rotation $SO(3)_{S}$ symmetry group is reduced
to
its $SO(2)_{S}=U(1)_{S_3}$ subgroup; while the breaking of the other two
symmetry groups in Eq.(2.1.1b) occurs in the same manner as in the electroweak
transition:
$$
SO(3)_{L}\times  U(1)_N \rightarrow
 U (1)_Q\equiv U(1)_{L_3-N/2} \ .
\eqno(2.1.2b)
$$
Here we have ignored a $Z_2$ factor that arises from the mixing of orbital
and spin rotations. We shall discuss this factor in Sec. 3.

Further we refer to the corresponding generator of the A-phase
symmetry,  which is  a generalized angular momentum operator
${\bf Q}={\bf L}_3-(1/2){\bf N}$, as the ``electric'' charge. It should be
emphasized that even though the scenario of the symmetry breaking in the two
systems have common features, the symmetry of the electroweak model is fully
gauged while that of $^3$He-A is fully global
\footnote{The ``semilocal''
limit of the electroweak model has also been used in studying
electroweak defects \cite{tvaa}. In this limit, the $U(1)_Y$ symmetry
is gauged but
the $SU(2)_T$ symmetry is taken to be global. This is similar to
the model of the anisotropic superconductivity in which $
(SO(3)_{S}\times  U(1)_N)/Z_2 \rightarrow U(1)_{S_z-N/2}$
\cite{BurlachkovKopnin}.
The electron liquid in superconductors is electrically
charged superfluid, as a result the  $U(1)_N$ symmetry is gauged, while the
$SO(3)_S$ symmetry is global.}.

The Higgs field (order parameter) in the electroweak model is the
spinor
$$
\Phi_{\rm ew}=\left(\matrix{\phi^+\cr \phi\cr}\right)~~,
\eqno(2.1.3a)
$$
which is normalized by:
$\Phi^{\dag} _{ew}
\Phi_{ew} = \eta_{ew}^2/2$ in the  vacuum where $\eta_{ew} \sim 250 GeV$.
The order parameter in superfluid $^3$He is a $3\times 3$ complex matrix
$A_{\alpha i}$ corresponding to the representation $(S=1, ~L=1)$ of the
rotation groups. (The 9 complex elements of this $3\times 3$ matrix transform
by a phase factor when acted by elements of the $U(1)_N$ group.)
Under the spin rotations the matrix transforms
as a vector in its Greek index,  while under the orbital rotations it
transforms  as a vector in its Latin index. In the A-phase vacuum of the
superfluid $^3$He, the order parameter can be factorized:
$$
A_{\alpha i}=\Delta_0 \  d_{\alpha } \Psi_i ~.
\eqno(2.1.3b)
$$
where, $\Delta_0 \sim 10^{-7} eV $ is the temperature dependent
amplitude of the gap in the quasiparticle spectrum (see below).
Here the spin part of the order parameter - the  unit vector
$\hat{\bf d}$ - denotes the axis of the spontaneous magnetic
anisotropy of the $^3$He-A vacuum.
The $^3$He-A counterpart of the electroweak Higgs field is the orbital
part of the order parameter and is written as a complex vector
$$
\vec  \Psi_{\rm 3 He-A}=
                        {{\hat e_1+i\hat e_2}\over \sqrt 2}~~,
\eqno(2.1.3c)
$$
with $\hat e_1\cdot \hat e_1=\hat e_2\cdot \hat e_2 =1$,
$ \hat e_1\cdot \hat e_2 =0$.

The orbital anispotropy axis
$$
\vec  l_{\rm 3 He-A}= i {{\vec  \Psi\times \vec  \Psi^\dagger} \over
                        {\vec \Psi^\dagger\cdot \vec  \Psi}}
                   = \hat e_1\times\hat e_2 ~~.
\eqno(2.1.4a)
$$
gives the  direction of the spontaneous orbital momentum
$<{\rm vac}\vert  {\bf L} \vert{\rm vac}>$ in $^3$He-A. It plays the part of
the
spontaneous isotopic spin $<{\rm vac}\vert  {\bf T}
\vert{\rm vac}>$ in the electroweak vacuum:
$$
\vec  l_{\rm ew}=-{  \Phi^\dagger\vec  \tau \Phi
                      \over {\Phi^\dagger  \Phi }} ~~,
\eqno(2.1.4b)
$$
where $\vec  \tau$ are the Pauli matrices in the isotopic space. (For
convenience we will drop the subscripts $ew$ and $^3$He-A on
$\Psi$ and $\vec \Psi$ in what follows.)

In $^3$He-A the superfluid  velocity of the vacuum flow is
$$
\vec  v_s=-{i\hbar\over 4m_3}
          {{(\Psi_i^\dagger\vec  \nabla \Psi_i -
                         \vec  \nabla\Psi_i^\dagger \Psi_i)} \over
          {\vec \Psi^\dagger\cdot \vec  \Psi}} ~~,
\eqno(2.1.5a)
$$
where $m_3$ the mass of the $^3$He atom. This velocity leads to the
superfluid mass current in the
$^3$He-A   vacuum (the mass current in the nonrelativistic condensed
matter coincides with the density of the linear momentum):
$$
\vec  j_N=m_3 n_N  \vec  v_s  ~~,
\eqno(2.1.6)
$$
where $n_N$ is the particle density of the $^3$He-A vacuum (the total $N$
charge of the $^3$He-A vacuum is the number of the $^3$He atoms: $N=\int dV
n_N$). In excited state or at nonzero temperature $T$ this current is
supplemented by the current of thermal fermions:
$$
\sum_{\vec p}\vec p f( \vec p)=\rho_n(\vec
v_n  -\vec  v_s)~,
\eqno(2.1.7)
$$
where
$$
f=(\exp [E_{\vec p}+\vec  v_s\cdot \vec p -\vec  v_n\cdot{\vec p}
]/T+1)^{-1}
$$
is the Fermi function;
$ E_{\vec p}$ is the spectrum of the single-particle fermionic excitations;
$\vec v_n$ is the velocity of the heat bath (usually denoted as the velocity
of
the normal  compoment  of the liquid) and $\rho_n$ is the density of the
normal
compoment ($\rho_n$   is a tensor in anisotropic $^3$He-A). The total mass
current of homogeneous superfluid is thus
$$
\vec  j_N=(m_3 n_N -\rho_n) \vec  v_s+ \rho_n \vec
v_n =  \rho_s \vec  v_s+ \rho_n \vec
v_n ~~, \eqno(2.1.8a)
$$
and the quantity $\rho_s=m_3 n_N -\rho_n$ is the so called (tensorial)
density
of the superfluid component of the liquid which appears below $T_c$ as the
manifestation of the breaking of $U(1)_N$ symmetry. Close to $T_c$
one has $\rho_s\propto m_3 n_N (\Delta_0(T)/T_c)^2$ with
$(\Delta_0(T)/T_c)^2\sim
 1-{T/ T_c}$.

In the electroweak vacuum the superfluid current corresponds to the
hypercharge current of the vacuum which appears as a result of the breaking of
$U(1)_Y$ symmetry
$$
\vec  j_Y=({\Phi^\dagger  \Phi })~  \vec  v_s  ~~,
\eqno(2.1.8b)
$$
with ${\Phi^\dagger  \Phi } \propto  1-{T\over T_c}$ near $T_c$.
The corresponding ``superfluid velocity''
$\vec  v_s$ is expressed in terms of the spinor order parameter
$$
\vec  v_s=
          {{ \Phi^\dagger\vec  D \Phi -
                       (\vec  D\Phi )^\dagger  \Phi }
          \over {2\Phi^\dagger  \Phi }} ~~,~~
\vec  D=-i\vec \nabla {\bf 1} -\vec  {\bf W} - \vec {\bf Y} ~~.
\eqno(2.1.9)
$$
where, $\vec {\bf W} = \tau_\alpha {\vec W}_\alpha /2$ and
$\vec {\bf Y} = {\bf 1} {\vec Y}/2$ are the $SU(2)_T$ and
$U(1)_Y$ gauge fields.
The difference between the electroweak model and $^3$He-A
is due to the gauge versus global symmetry groups of the two systems: The
equation (2.1.9) for the electroweak current contains the gauge field
$Z=T_3-Y/2$.

In the electroweak case one can define 4 conserved currents
corresponding to the 4 generators of the symmetry group $SU(2)_{T}\times
U(1)_Y$. In the case of normal $^3$He there are 7 generators of the
symmetry group
$ SO(3)_{S}\times SO(3)_{L}\times  U(1)_N $, but only 4 currents
are conserved and have physical
meaning. These are: the mass current related to $N$ (which corresponds to
the hypercharge current) and 3 spin currents related to the
$ SO(3)_{S}$ generators. The ``internal''  orbital momentum $\vec  L$ of the
Cooper pairs is not a conserved quantity, because it can be transferred to the
``external'' orbital momentum
$\vec  r\times \vec  j_N$ associated with the macroscopic flow. That is why
after the symmetry breaking into the
$^3$He-A state, on the {\bf Level 1} there is no analog to the (conserved)
current of the  little group (electromagnetism). (Interaction between the
internal and external orbital momenta is important for the vortex structure
in $^3$He-A (see Eq.(3.1.8)).

When the orbital anisotropy axis, $\hat {\bf l}$, is constant, the superfluid
velocity $\vec  v_s$ is the gradient of the phase of the order parameter and
thus is curl-free.  If, however, $\hat {\bf l}$ varies in space, forming  the
so-called
$\hat {\bf l}$-texture, the $^3$He-A mass flow acquires vorticity, which is
expressed
in terms of the $\hat {\bf l}$ field  by the Mermin-Ho relation
\cite{Mermin-Ho}:
$$
\vec  \nabla\times\vec  v_s =
      {\hbar\over 4m_3~}e_{ijk} \hat l_i\vec  \nabla \hat l_j\times\vec
\nabla
\hat l_k ~~.
\eqno(2.1.10)
$$
This follows from Eq.(2.1.5a).

\subsection{Bosonic sector of the Electroweak Model} \label{Bosonic/el}

The standard model of the electroweak interactions
is an $SU(2)\times U(1)$ invariant theory with a scalar field
$\Phi$ in the fundamental representation of $SU(2)$. The bosonic
sector of the model is described by the Lagrangian:
$$
L_b = L_W + L_Y + L_{\Phi} - V(\Phi_{\rm ew} )
\eqno(2.2.1)
$$
where,
$$
L_W = - {1 \over 4 g^2} W_{\mu \nu a} W^{\mu \nu a}
\eqno(2.2.2)
$$
$$
L_Y = - {1 \over 4 {g'}^2} Y_{ \mu \nu} Y^{\mu \nu}
\eqno(2.2.3)
$$
where $W_{\mu \nu}^a$ and $Y_{\mu \nu}$ are the field
strengths for the $SU(2)_T$ and $U(1)_Y$ gauge fields $W_\mu ^a$ and $Y_\mu$
respectively. Also,
$$
L_\Phi = |D_\lambda \Phi_{\rm ew} |^2 \equiv
   \biggl |\biggl (\partial _\lambda -
             {{i } \over 2} \tau ^a W_\lambda ^a -
                  {{i } \over 2} Y_\lambda \biggr)\Phi_{\rm ew} \biggr| ^2
\eqno(2.2.4)
$$
$$
V(\Phi_{\rm ew} ) = \lambda (\Phi_{\rm ew} ^{\dag} \Phi_{\rm
ew} - \eta_{\rm
ew} ^2 /2 )^2 \ ,
\eqno(2.2.5)
$$
where,
$$
\Phi_{\rm ew}=\left(\matrix{\phi^+\cr \phi\cr}\right)~~,
\eqno(2.1.3a)
$$
is a complex doublet.

Below the symmetry breaking transition the transverse fields ${W_\mu}^1$,
${W_\mu} ^2$ and the following combination of the longitudinal ${W_\mu}^3$
field and ${Y_\mu}$
$$
Z_\mu \equiv  l_{ew}^a {W_\mu} ^a -  Y_\mu \
\eqno(2.2.6)
$$
acquire masses, while the combination
$$
A_\mu \equiv  \sin^2\theta_W l_{ew}^a {W_\mu} ^a + \cos^2\theta_W  Y_\mu
\ ,
\eqno(2.2.7)
$$
remains massless. This massless gauge field represents the little group
$U(1)_Q$ of electromagnetism. Here
$\hat{\bf l}_{ew}$ is the unit vector defined in Eq. (2.1.4b); the weak
mixing angle $\theta_W$ and electric charge $e$  are  given by the
equations $e = g\sin \theta_W = g' cos\theta_W$.

\

\subsection{Bosonic sector of $^3$He} \label{Bosonic/3He}

In superfluid $^3$He the bosonic  sector consists of two groups: (i) The
soft variables, which are the densities of the conserved quantities.
They exist even in normal state above transition. These are the  particle
density $n_N$, the spin density $\vec  S$ and the density of the linear
momentum $\vec j_N$. The latter is usually expressed in terms of the velocity
of the liquid: in the normal liquid above transition $\vec j_N=m_3n_N\vec v$;
in superfluid state this variable transforms to the normal velocity
$\vec  v_n$ of the heat bath of  fermionic (and/or  bosonic) excitations in
Eq.(2.1.8a). (ii) The order parameter (Higgs) field $A_{\alpha i}$  which
appears below transition. In general the dynamics of all these variables is not
described by a Lagrangian because of dissipation and the nontrivial
interaction of these variables with the fermionic degrees of freedom. There
are only a few regimes where a simple Lagrangian description is possible:

\noindent (1) Close to $T_c$  there is the  Ginzburg-Landau free
energy functional, which describes the static Higgs fields $A_{\alpha
i}$. In some cases the  Ginzburg-Landau description can be extended to include
the time dependence.

\noindent (2) Hydrodynamic description  in terms of only the soft
variables ($n_N$,  $\vec  S$, $\vec  v_n$ + Goldstone bosons sector of the
Higgs fields)
works in all  temperature regimes  if the dynamics is slow enough
as compared to the time of establishment of the local thermal equilibrium.

\noindent (3) The quantum field theory at near zero temperature,
where the dynamics of
the soft bosons and low energy fermions can be constructed.

We will now discuss each of these descriptions in some detail.

The {\it Ginzburg-Landau free-energy functional}
must be invariant under the total
symmetry group $G$  of the physical laws. This invariance essentially restricts
the number of fourth-order terms and also the number of gradient terms that
can be written. In superfluid
$^3$He it contains one  second-order term,  five fourth-order terms and three
gradient terms.  In each of them the Greek spin indices  should not  be  mixed
with the Latin orbital indices to provide the invariance under
separate spin $SO(3)_{(S)}$ and orbital $SO(3)_{(L)}$ rotations;
also each term should contain an equal number of
$ A^{\ast }_{\alpha i}$ and $ A_{\alpha i}$ for invariance under $U(1)_N$
transformations. As a result of these requirements
$F^{G-L} \lbrack A_{\alpha i}\rbrack $ is given by \cite{VolWol}
$$F^{G-L}  = -\alpha A^{\ast }_{\alpha i} A_{\alpha i}
+\beta _1A^{\ast } _{\alpha i} A^{\ast } _{\alpha i}
A_{\beta j} A_{\beta j}
+\beta _2A^{\ast }_{\alpha i}
A_{\alpha i} A^{\ast } _{\beta j} A_{\beta j}
+\beta _3 A^{\ast } _{\alpha i} A^{\ast } _{\beta i}
A_{\alpha j} A_{\beta j}$$
$$+\beta _4 A^{\ast }_{\alpha i} A_{\beta i} A^{\ast }_{\beta j}
A_{\alpha j}
+\beta _5 A^{\ast }_{\alpha i} A_{\beta i} A_{\beta j} A^{\ast}_{\alpha j}$$
$$+ \gamma_1 D_j^{\ast } A^{\ast }_{\alpha i}D_j A_{\alpha i}+
  \gamma_2 D_j^{\ast } A^{\ast }_{\alpha i}D_i A_{\alpha j}
+ \gamma_3 D_j^{\ast } A^{\ast }_{\alpha j}D_i A_{\alpha i}
\hskip2mm . \eqno(2.3.1)$$
Here $\vec  D= -i\vec \nabla  -m_3\vec  v_n$, where the velocity of the normal
component is fixed; the parameter
$\alpha$ changes sign at
$T_c$,    $\alpha =\alpha_0(1-T/T_c)$, while $\alpha_0$ and the $\beta$'s are
functions of pressure only and  depend on the details of the microscopic
interaction of the  $^3$He atoms. Note that  the last two gradient
terms are not invariant under separate (isotopic) orbital rotation of the order
parameter: this results from the interaction of the internal and external
orbital rotations.

The vacuum manifold resulting from the minimization of the
Ginzburg-Landau functional depends on the parameter values entering
eq. (2.3.1) and hence on the temperature and pressure. Therefore, the
values of the temperature and pressure determine the superfluid phase.

The Ginzburg-Landau functional is also useful for determining the core
structure of singular topological defects, within which the order
parameter deviates from its vacuum values. The core size is of the order of
coherence length
$$
\xi(T)\sim \sqrt{\gamma/\alpha}\sim (200-500) (1-{T/T_c})^{-1/2}A^o\ .
$$
This correspondes to the scale $\xi(T)=1/m_{Higgs}(T)$
determined by the inverse mass of the Higgs boson in electroweak theory,
which defines the core size of $Z$ string.

\

The {\it hydrodynamic or London energy} is the energy of fields on
the vacuum manifold of a given superfluid phase.  In $^3$He-A the London
energy
is given in terms of the mass density $\rho=m_3n_N$, spin density $\vec  S$,
velocity $\vec  v_n$ of the normal component,
orbital  Goldstone variables $\vec  v_s$,  $\hat{\bf l} $
and  spin Goldstone field $\hat{\bf d}$:
$$
F^{London}  =F(\rho,T) + {1\over 2} \gamma ^2 S_{\alpha } (\chi ^{-1}) _
{\alpha \beta } S_{\beta } -\gamma \vec  B \cdot \vec  S+ {1\over 2}
(\rho\delta
_{ij}-(\rho_s) _{ij}) (\vec  v_n)_i (\vec  v_n)_j
$$
$$
+ {1\over 2} (\rho_s) _{ij} (\vec  v_s)_i (\vec  v_s)_j  + {1
\over 2}K_{ijmn}\partial_i\hat{\bf l}_m \partial_j\hat{\bf l}_n  + C_{ij}
(\vec  v_s)_i
(\vec  \nabla \times \hat{\bf l} \, \, )_j
$$
$$
+{1\over 2} (\rho_{sp}) _{ij} \nabla_i
\hat{\bf d} _ {\alpha }\nabla_j \hat{\bf d} _ {\alpha }- {1\over 2}g_
{so}(\hat{\bf d}
\cdot\hat{\bf l})^2
\hskip2mm .
\eqno(2.3.2)
$$

Here $\vec  v_s$ is defined with the factor $\hbar/2m_3$ multiplied into
Eq.(2.1.5a); the tensors  $(\rho_s) _{ij}$, $C _{ij}$,
$(\rho_{sp}) _{ij}$ and $K_{ijmn}$ are
uniaxial tensors with the anisotropy axis along $\hat{\bf l}$:
$$
(\rho_s) _{ij} = \rho ^{\parallel } _s \hat{\bf l}_i\hat{\bf l}_j + \rho
^{\perp }_s
(\delta _{ij} - \hat{\bf l}_i\hat{\bf l}_j) \hskip2mm ,
\eqno(2.3.3)
$$
$$
C _{ij} =  C\delta _{ij}-C_0 \hat{\bf l}_i\hat{\bf l}_j   \hskip2mm ,
\eqno(2.3.4)
$$
$$
{1 \over 2}K_{ijmn}\partial_i\hat {\bf l}_m \partial_j\hat {\bf l}_n = {1
\over 2}  \Bigl
\lbrack
K_1 (\vec \nabla \cdot \hat {\bf l} \, \, )^2
+ K_2 (\hat {\bf l} \cdot (\vec \nabla
\times \hat {\bf l} \, \, ))^2 +K_3 (\hat {\bf l} \times (\vec \nabla
\times \hat {\bf l} \,
\, ))^2
\Bigr \rbrack  \hskip2mm .
\eqno(2.3.5)
$$
Here $K_1$, $K_2$, and
$K_3$ are the twist, splay, and bend coefficients respectively, and,
$\chi_{\alpha \beta }$ is a  uniaxial tensor
with the anisotropy axis along $\hat{\bf d}$:
$$
\chi _{\alpha \beta} = \chi _{\parallel }
\hat{\bf d}_{\alpha}\hat{\bf d}_{\beta} + \chi _{\perp }
(\delta _{\alpha \beta} - \hat{\bf d}_{\alpha}\hat{\bf d}_{\beta})
\hskip2mm .
\eqno(2.3.6)
$$
$\vec  B$ is an external magnetic field interacting with the spin  density. The
last term in Eq.(2.3.2) describes the tiny spin-orbital  coupling  between
$\hat{\bf d}$  and $\hat{\bf l}$.

This functional is  useful for determination of the continuous
structures of the Goldstone fields (textures) in which the system
does not leave the vacuum manifold.

The mass current in the London limit is
$$
m_3(\vec  j_N)_i ={\partial F^{London}\over \partial v_{si} }+
{\partial F^{London}\over \partial v_{ni} }
$$
$$
= (\rho_s) _{ij} (\vec  v_s )_j + (\rho\delta
_{ij}-(\rho_s) _{ij}) (\vec  v_n )_j+ C _{ij}
(\vec  \nabla \times \hat{\bf l})_j
\hskip2mm             .
\eqno(2.3.7)
$$
The current contains three terms of which the first and second are
due to the superfluid and normal component motion and
the third is due to the $\hat{\bf l}$ texture.

\

The {\it quantum description} includes the dynamics of  Goldstone fields
and the propagation of the elementary particles (fermions and bosons) in the
potentials produced by these  Goldstone fields. Typical example is the
interaction of the fermionic particle with the field of quantized vortex.

\section{Strings} \label{strings}

\subsection{  Singular Strings}

The {\bf Level 1} analogy between the electroweak model and $^3$He
leads to similarity in the structure of defects in the
two models, though there are important differences too.
The main difference is that the topology of the vacuum manifold
in the electroweak model does not support
topologically stable strings while the vacuum manifold for
$^3$He-A admits topological strings in addition to analogs of
electroweak strings. The fundamental homotopy group of
the electroweak vacuum manifold
$$
M_{\rm ew} =G_{\rm ew}/H_{\rm ew}=SU(2)~~
\eqno(3.1.1)
$$
is trivial: $\pi_1(M_{\rm ew})=0$.

In contrast, the A-phase manifold
$$
M_{\rm A} =G_{\rm 3He}/H_{\rm A}=(SO(3)\times S^2)/Z_2~~
\eqno(3.1.2)
$$
has nontrivial fundamental  group
$$
\pi_1(M_A)=Z_4~~,
\eqno(3.1.3)
$$
which contains four elements. There are two reasons for this: (i) instead of
$SU(2)$ in $M_{\rm ew}$ the A-phase manifold $M_{\rm A}$ contains the group
$SO(3)=SU(2)/Z_2$, and, (ii) the full symmetry group of the A-phase vacuum:
$$
H_{\rm A}= U(1)_{S_3}\times U(1)_{L_3-N/2}\times Z_2~~.
\eqno(3.1.4)
$$
contains another $Z_2$ factor. This is the symmetry under
rotation of the spin axis, $\hat{\bf d}$, by $\pi$ about an axis perpendicular
to $\hat{\bf d}$ followed by a
rotation of the triad ($ \hat e_1$, $ \hat e_2$, $ \hat{\bf l}$) by $\pi$ about
$\hat{\bf l}$ in orbital space. According to Eq.(2.1.3b), this leaves
the full order parameter invariant.
Both discrete symmetries $Z_2$, being combined, give the
$Z_4$ in Eq.(3.1.3).

This means that there are 4 topologically distinct classes of strings in
$^3$He-A. Each
can be described by the  topological charge $\nu$ which takes only 4 values,
which we choose to be $0$, $\pm~1/2$ and $1$ with summation modulo 2 (ie.
1+1=0).

Let us now construct different strings and distribute them into classes
characterized by the charge $\nu$. The strings - corresponding to $U(1)$
vortices with integer winding number $n$ - have the following asymptotic
form:
$$
A_{\alpha j}( r \rightarrow \infty,\phi)=
\Delta_0 \ e^{in\phi}\hat z_{\alpha }(\hat x_j+i \hat y_j) ~~.
\eqno(3.1.5a)
$$
The circulation of the superfluid velocity around the vortex core is
quantized
$$
\oint d\vec  r \cdot\vec  v_s = \kappa n~~,~~\kappa={2\pi\hbar\over 2m_3}~~,
\eqno(3.1.6)
$$
and $\kappa$ is called the circulation quantum.
The order parameter phase
$\phi$ is not defined on the vortex axis,
i.e. the vorticity is singular on the
vortex axis.

The asymptotic form of vortices with fractional circulation number
$n=\pm~1/2$ (or, simply, half-quantum vortices) is given by
$$
A_{\alpha j}( r \rightarrow \infty,\phi)=\Delta_0 \ e^{\pm i\phi/2}
(\hat x_{\alpha }\cos{\phi\over 2} +\hat y_{\alpha }\sin{\phi\over 2})(\hat
x_j+i \hat y_j)~~.
\eqno(3.1.5b)
$$
On circumnavigating such a vortex, the change of the sign of the order
parameter due to the phase winding by $\pi$, is compensated by the change of
sign of the vector
$\vec {\bf d}=\hat {\bf x}\cos{\phi\over 2} +\hat {\bf y}\sin{\phi\over 2}$.
This vortex is the counterpart of Alice strings considered in particle
physics\cite{Schwarz}: a particle that goes around an Alice string flips its
charge. In
$^3$He-A the quasiparticle going around a $1/2$ vortex flips its
$U(1)_{S_3}$ charge, that is, its spin.
As a consequence, several phenomenon ({\it eg.} global Aharanov-Bohm
effect) discussed in the particle
physics literature have corresponding discussions in the condensed matter
literature (see \cite{Khazan},\cite{SalVol} for $^3$He-A and
\cite{Russel} in particle physics).  Note   that in  $^3$He-A also the
particle-like topological object, the hedgehog, flips its $\pi_2$ topological
charge around the 1/2 vortex \cite{VolMin} .

All the vortices in $^3$He-A can now be grouped in accordance with their
$\pi_1$
topological charge $\nu$. The  half-quantum vortices
belong to the classes
$\nu=\pm~1/2$; the  vortices with odd circulation number,
$n=2k+1$, belong to the class $\nu=1$, and the vortices with even
circulation
number, $n=2k$, have
zero topological charge $\nu=0$. The latter means that, as distinct
from lines
with $\nu=\pm~1/2,~1$, the singularity in these $\nu=0$ strings, such as
$n=2$
vortex, can be continuously dissolved. What is left is called
``texture''  with continuously distributed (non-singular)
order parameter within the vacuum manifold.

The $n=2$ ($\nu =0$) topologically unstable vortex corresponds to the unit
winding $Z$ string in the electroweak model. Both the $Z-$string and
the $^3$He-A $n=2$ vortex have constant
$\vec  l=\hat z$, and, in both cases the phase of the order parameter
has a singularity at $r=0$. The Higgs field configuration
for a $Z-$string is:
$$
\Phi_{\rm Z-string}(\vec  r) =
    {\eta \over {\sqrt{2}}} ~f (r)~ e^{i\phi}
    \left(\matrix{ 0\cr 1\cr}\right)~~,
\eqno  (3.1.7)
$$
in cylindrical coordinates. In the electroweak case, the form of the
Higgs field alone does
not describe a string and the gauge fields must also be specified. The only
non-vanishing gauge field in the $Z-$string is $Z_\phi = -2 v(r)/(\alpha r)$
where $v(0)=0$ and $v(\infty )=1$. Like in superconductors, due to
Meissner effect the gauge field $Z$ screens the superfluid velocity
far from the core,
where $\oint d\vec  r \cdot \vec  v_s \rightarrow 0$. This is different
from the electrically neutral $^3$He where this complication is absent since
there are no gauge fields and circulation is conserved quantity (if $\hat
{\bf l}=\hat z$ is fixed).

The singularity at the origin is smoothed out by
escaping the Higgs field from its vacuum manifold with $f=1$. In the
axisymmetric $Z$-string one has $f(0) = 0$ and
$f(r\gg \xi(T))=1$ at radial infinity (Fig.1a). In $^3$He-A the behavior
within the core of the size $\sim\xi(T)$ is slightly different. The structure
of
the axially symmetric Higgs field in the A-phase vortex core is described by
two radial functions consistent with the axial symmetry (Fig.1b):
$$
A_{\alpha j}(r,\phi)=
\Delta_0 \hat z_{\alpha }[ \ e^{in\phi}  f_1(r)(\hat x_j+i  \hat
y_j)  +\ e^{i(n+2)\phi}  f_2(r)(\hat x_j-i  \hat y_j)]~~.
\eqno(3.1.5c)
$$
Here $f_1(r\gg \xi(T))=1$, $f_2(r\gg \xi(T))=0$, $f_1(0)=0$, while
$f_2(0)$ depends on $n$: at $n=-2$ the parameter $f_2(0)$ is finite,
i.e. the Higgs field does not necessarily disappear in the vortex core.

The reason for the appearance of the
$f_2(r)$ term as compared with the electroweak string is the interaction
of the
internal (isotopic) orbital rotations
$SO(3)_L$  of the vectors $\hat e_1$ and $\hat e_2$ with the coordinate
rotations. The order parameter in Eq. (3.1.5c) is the eigenstate of the total
angular momentum  ${\bf L}_3^{\rm total}$
$$
{\bf L}_3^{\rm total}={\bf L}_3^{\rm int} + {1\over i} \partial_\phi~~,
\eqno(3.1.8)
$$
with $L_3^{\rm total}=n+1$. This momentum is distributed between the
internal and external momenta in the following way: in the first term
$L_3^{\rm int}=1$ , $L_3^{\rm ext}=n$, while in the second one
$L_3^{\rm int}=-1$ , $L_3^{\rm ext}=n+2$. The second term represents the
component of the Higgs field with the opposite orientation of the
$\hat{\bf l}$, which appears in the vortex core. (Also see Sec.(6.1).)
In the electroweak string such component can appear
only by the additional spontaneous symmetry breaking in the vortex core
due to the instability of Z-string induced, say,  by the fermion zero modes
\cite{Naculich}.

The $n=2$ singular line in $^3$He-A and $Z$-string are topologically
unstable, since even with given asymptote at $r\rightarrow\infty$ the
singular core where the Higgs field deviates from its vacuum value can
be removed by the formation of the continuous
$\hat {\bf l}$ texture with the vacuum manifold
everywhere in space (Anderson-Toulouse-Chechetkin
vortex,\cite{Anderson-Toulouse}, see Sec.3.2). That is, in
$^3$He-A {\bf \cite{Blaha}} and in the electroweak model, the
vortices can
terminate on a $\hat {\bf l}$ hedgehog ($\hat {\bf l}=\hat r$);
in particle
physics this hedgehog configuration is a {\it magnetic monopole}
of the 't Hooft-Polyakov type with an additional
physical string \cite{nambu}.

In the electroweak model, we also have the $W-$string solutions.
They correspond  to the so called disgyrations in $^3$He-A: the lines
with winding of the $\hat{\bf l}$-vector without winding of the phase.
For example two (gauge equivalent) $W-$string configurations are:
$$
\Phi^{(1)}  (r,\phi )=
{\eta \over {\sqrt{2}}}
f_{W}(r)\pmatrix{\cos \phi \cr i\sin \phi \cr} \ ,
$$
$$
{\bf W}^1={v_{W}(r)\over r}\hat {\bf e}_\phi ,~
,\ \ {\bf W}^2= {\bf W}^3={\bf B}=0
\eqno (3.1.9)
$$
$$
\Phi^{(2)} (r,\phi) = {\eta \over 2}f_{W}(r)
\pmatrix{e^{i\phi }\cr e^{-i\phi }\cr} ,
$$
$$
{\bf B}={\bf W}^1= {\bf W}^2=0,\ \ {\bf W}^3=
{v_{W}(r)\over r}\hat {\bf e}_\phi ,~
\eqno (3.1.10)
$$
where, the subscript $W$ on the profile functions $f$ and $v$
means that the functions are the ones appropriate to the
$W$-string. In $^3$He, the $W$-strings
correspond to the topologically unstable disgyrations with
$n_l=2$ winding number for the $\hat{\bf l}$-vector around the
line:
$$
\hat {\bf l}^{(1)}=\hat z\cos 2\phi +
\hat y\sin 2\phi~~,~~
\hat{\bf l}^{(2)}=\hat x\cos 2\phi - \hat y\sin 2\phi \ .
\eqno (3.1.11)
$$

Thus we see that the similar symmetry groups of $^3$He and the
electroweak model lead to similar vortex and monopole structures
in the two systems.
The different nature of the symmetries - global versus gauge -
does not play
a role in the topology of such textures in the two systems, though
it does influence the energetics of these defects.

\subsection{Continuous Textures}

In $^3$He-A, the ``textures'' - spatially inhomogenous distributions in  the
vector fields $\hat {\bf d}$ and $\hat {\bf l}$ - are similar to those in
liquid
crystals and in ferro- and antiferromagnets. (Note that in cosmology a texture
has a more narrow meaning: it is a point-like object  - Skyrmion - with a
continuous distribution of the order parameter described by the topological
charge of the homotopy group $\pi_3(M)$). The spatial scales of these
textures are defined either by the dimension of the vessel or by some fine
interaction like tiny spin-orbit coupling $-(1/2) g_{so} (\hat{\bf d}\cdot\vec
l )^2$ in Eq.(2.3.2)
and interaction $\hat{\bf d}$ with external magnetic field.
These scales are essentally larger then the coherence length $\xi(T)$ which
defines the core of singular defects.

A domain wall in soft ferromagnets  is one of the numerous examples of textures
which scale   essentially exceeds $\xi(T)$. This texture exists due to small
spin-orbit coupling of magnetization $\vec S$ with the crystallographic
direction of the  easy axis $-(1/2) g_{so} (\hat z\cdot\vec S )^2$. This
interaction reduces the  $M=S^2$ vacuum manifold of the isotropic ferromagnet
to the the manifold $\tilde M=Z_2$, which contains 2 disconnected pieces $\vec
S=\pm S\hat z$.  Stability of the domain wall is guaranteed by the nontrivial
group
$\pi_0(\tilde M)=Z_2$:  the domain wall in ferromagnets can be either closed
or terminate on the boundary of the system but it cannot have a boundary.

In $^3$He-A the situation is more complicated: the vacuum manifold $M_A$
contains
only one connected piece, even when it is reduced by the spin-orbit
interaction. In the presence of the spin-orbit coupling the reduced
vacuum manifold is $\tilde M_A=\tilde G/\tilde H_A$. Here the initial symmetry
group $\tilde G= SO(3)^{J}\times U(1)$ contains only simultaneous spin and
orbital rotations and the invariant subgroup is
$\tilde H_A= SO(2)^{J}\times U(1)$). As a result $\pi_0(\tilde M_A)$ is
trivial, and does not support topological domain walls. Nevertheless, one can
construct a continuous  planar texture, separating two volumes with different
orientation of
$\hat {\bf l}$ with respect to  $\hat {\bf d}$ (see Fig.~2). This
looks like a domain wall in ferromagnets, but the topology which supports its
stability is different.  This   wall can terminate and   the edge of
the wall is the location of a topologically stable singular line:  a
1/2-quantum vortex. Therefore this wall cannot be destroyed in a
non-singular way - that is, the order parameter must leave the
vacuum manifold $M_A$ if the wall is to terminate - and thus it also has a
topological charge.  As distinct from  walls with nontrivial $\pi_0$ we call
this  object a topological {\it soliton}. Such solitons were identified in NMR
experiments on  $^3$He-A. (See Reference \cite{Crossing} where
the crossing of the soliton with the nonsingular
Anderson-Toulouse-Chechetkin vortices has been observed.
The crossing point represents the ``texture'' in
the particle physics sense: like a Skyrmion it is described by the $\pi_3$
topology \cite{TopologyOfCrossing}).

The stability of the soliton is dictated by the same homotopy group,
which is responsible for the stability of walls bounded by strings,
discussed in cosmology \cite{Vilenkin}.
This is the  relative homotopy group $\pi_1 (M_A,\tilde M_A)$,
which deals with different vacuum manifolds at different scales: far from the
soliton at a distance larger than the characteristic length $\xi_D$ of
spin-orbit coupling the  vectors  $\hat {\bf d}$ and $\hat {\bf l}$ are locked
together  (we call this as a dipole locking)
and the vacuum manifold is reduced to $\tilde M_A$, while within the soft
core of the soliton of size $\xi_D$ it is again restored to $ M_A$.

\subsection{Topology of Vortex Textures}\label{s.Topology}

The axisymmetric $\hat {\bf l}$ texture is defined by two radially
dependent functions:
$$
\hat {\bf l} = \hat z \cos \eta (r) +
\sin\eta (r) ( \hat r \cos \alpha (r)+ \hat \phi \sin \alpha (r)) ~~.
\eqno(3.3.1)
$$
This is a general  solution of the equation of the axisymmetry for the orbital
texture:
$$
{\bf L}_3^{\rm total}~ \hat {\bf l} (\vec r)=0~~,
\eqno(3.3.2)
$$
where ${\bf L}_3^{\rm total}$ is the generator of orbital rotations in
eqn.(3.1.8).

The singular $n=2$ vortex (and also the electroweak $Z-$string)
corresponds
to
$\eta (r)=0$ for all $r$. But there is another vortex - called the
Anderson-Toulouse-Chechetkin $4\pi$ vortex - which is non-singular and
which can be obtained by continuous deformations of the $n=2$ vortex.
This corresponds to having $\eta (r=0) =0$ and $\eta (r \ge r_0 ) = \pi$. The
winding number of this vortex outside the core is:
$$
n={1\over \kappa}\oint_{r>r_0} d\vec r \cdot \vec v_s
={1\over \kappa}\int d\vec S \cdot \vec \nabla \times \vec v_s =
{1\over 2\pi}\int dx~dy~\hat{\bf l}\cdot (\partial_x \hat{\bf l}
\times \partial_y\hat{\bf l})=2 ~~,
\eqno(3.3.3)
$$
Here we used the Mermin-Ho relation and the expression for the topological
invariant which describes the mapping $S^2\rightarrow S^2$ of the vortex
cross-section to the sphere $S^2$ of the unit vector
$\hat{\bf l}\cdot\hat{\bf l}=1$. The invariant shows that within the
continuous $4\pi$-vortex the full area ($4\pi$) of the sphere is swept once.

The vortex core radius $r_0$ usually essentially exceeds $\xi(T)$: it
can be limited by the next scale in the hierarchy of interactions,
which can be
the radius of the vessel or the scale $\xi_D$ of the spin-orbit coupling. We
call this the soft core compared to the hard  core of the singular defect.

Thus the  $n=2$ vortex escaped the singularity but immediate pay for it is
the  broken parity: in the texture formed one has $\hat{\bf l}(\vec r)\neq
\hat{\bf l}(-\vec r)$, though some combined parity is still retained.
This is one of the numerous examples in $^3$He, when the reduction of energy
is accompanied by parity violation.

Such continuous  vortices are described by $\pi_2$ homotopy. If
$\hat{\bf l}$ is fixed at infinity we can define two charges of the
vortex
$$
\nu_l= {1\over 4\pi}\int dx~dy~\hat{\bf l}\cdot (\partial_x \hat{\bf l}
\times \partial_y \hat{\bf l})  ~~,
\eqno(3.3.3)
$$
and
$$
\nu_d= {1\over 4\pi}\int dx~dy~\hat{\bf d}\cdot (\partial_x \hat{\bf d}
\times \partial_y \hat{\bf d})  ~~,
\eqno(3.3.4)
$$
These charges characterize the orientational  distribution of the
unit vectors $\hat {\bf l}$ and $\hat {\bf d}$. For the
Anderson-Toulouse-Chechetkin vortex
$\nu_l=n/2=1$, while $\nu_d$ can be either 1 or 0, depending on the
(experimental) external conditions (see \cite{PhaseDiagram} and Fig.3).

In the particle physics literature, Preskill's semilocal skyrmion
is analogous to the Anderson-Toulouse-Chechetkin vortex. In \cite{preskill}
the continuous deformation of a semilocal string (a $Z-$string in the
limit that $SU(2)_T$ is global \cite{tvaa}) into a non-singular
configuration is described. The final configuration is called a semilocal
Skyrmion and  corresponds directly to the global
Anderson-Toulouse-Chechetkin vortex in $^3$He-A. The semilocal
Anderson-Toulouse-Chechetkin vortices in one of the models of the anisotropic
superconductivity was considered in \cite{BurlachkovKopnin}.

For a periodic vortex array in a rotating vessel these topological charges
characterize the fields within an elementary cell of the vortex lattice.
The vortex cell with $n=2$ and $n=4$ circulations along the cell boundary
are most probable; Fig.~3b shows an elementary cell of the periodic texture
with $n=4$ circulation along the cell boundary. The situation for the
$\hat {\bf d}$ texture is more diverse. Without magnetic field
$\hat {\bf d}$ is dipole-locked with $\hat {\bf l}$, which means that the
$\hat {\bf d}$ texture has the same
$\pi_2$ charge as the $\hat {\bf l}$ texture
and so: $\nu_d=1$  for the texture in Fig.~3a and $\nu_d=2$ for the texture
in Fig.~3b.

The situation changes in an external magnetic field, where a new
length scale appears.
If we subject the texture to an axially oriented external
magnetic field and start increasing its strength, the coupling of the magnetic
part of the order parameter ($\hat {\bf d}$) to the field soon wins over the
spin-orbit interaction and $\hat {\bf d}$ will become confined to the plane
transverse to the magnetic field. The topology of the orbital $\hat {\bf l}$
field, which is responsible for the continuous vorticity, is insensitive to
this
process and thus $\nu_l = 1$, independently of the value of the field. However,
the magnetic topological number $\nu_d$ will undergo a discontinuous change
from
1 to 0. The final configuration is known  as the ``dipole-unlocked vortex''. In
the center of a dipole-unlocked vortex a continuous (non-singular) {\it
``soft''
vortex core} with a diameter on the order of $\xi_D$ is formed in which $\hat
{\bf d}$ is  dipole-unlocked  from $\hat {\bf l}$.  The change in the
configuration of the $\hat {\bf d}$ texture as a function of magnetic field
corresponds to a first order textural phase transition in the vortex structure
(see below).

\subsection{Observation and Transition of Vortices}\label{s.Transition}

Until now three types of quantized vortex lines \cite{PhaseDiagram} and the
vortex sheet (the soliton with accumulated vorticity \cite{VortexSheet},
see Fig.~3c) have been experimentally verified to exist in $^3$He-A which is
placed in a rotating container with an axial, external magnetic field
(Fig.~4 and Ref.\cite{PhaseDiagram}). The  three vortices are: 1) The
dipole-locked continuous vortex, which have
the following set  of topological charges
$\nu=0$, $\nu_l=n/2=1$, $\nu_d=1$ (Fig.~3a).
2)  The dipole-unlocked continuous vortex with
$\nu=0$, $\nu_l=n/2=1$, $\nu_d=0$ (Fig~.3a).
3) The singly quantized ($n=1$)
isolated vortex, which is singular since it has  a nonzero $\pi_1$ charge
($\nu=1$).

The structure of the latter vortex in Fig.~5 is rather peculiar: it
has a hard (singular) core of size $\sim \xi(T)$, but in the vicinity of the
core it represents not the vortex but  the disgyration  -- the singular line in
the $\hat{\bf l}$ field. It is analogous to the $W$-string in eqn.(3.1.11)
but with unit  winding number $n_l=1$. This singular line has no singularity
in the vorticity: the vorticity is thus continuous and is concentrated in a
soft core of dimension $\xi_D$, in which the $\hat {\bf l}$ texture has the
fractional $\pi_2$ charge $\nu_l=1/2$, corresponding to the continuous
vorticity of $n=1$ vortex. Within
the soft core the $\hat {\bf l}$ texture is dipole unlocked from the
$\hat {\bf d}$ texture, which has  $\nu_d=0$. This dipole unlocked structure
allows the observation of  the $n=1$ vortex in NMR experiments
\cite{PhaseDiagram}.

In the phase plane formed by the experimentally controlled variables
$\Omega$ (angular velocity) and $H$ (magnetic field strength)
each of the three vortex arrays  occupies a region
where its energy is  less than those of the others (Fig. 4a). These
regions are separated from each other  by first order phase transitions.
This does not
necessarily mean that in a given region of the $\Omega - H$ phase diagram we
will always observe the equilibrium (least energy) vortex array. Since the
transitions are of first order, a large energy barrier may separate two
different vortex structures. A vortex may thus remain
quite comfortably in a metastable state, even if the external variables are
changed while it already exists and it is transported into a foreign territory
of the phase diagram, where it is not the equilibrium structure.
For example, the vortex sheet state of the rotating $^3$He has been observed
only as a metastable state. The region where the vortex sheet state
has the lowest energy in  the rotating container
nevertheless exists (see Ref.\cite{PhaseDiagram} and Fig.~4b) but at higher
rotation velocity.

Especially large energy barrier exists for the transition between
the singular and continuous vortices.
This process is possible from the topological point of
view: the arithmetic summation laws for the charges $n$ and $\nu$
($1+1=2$ and  $1+1=0$ corespondingly) show that the result from a
fusion of two singular vortices is one continuous vortex.
However such a process has not been
observed because of the strong ``Coloumb'' repulsion of two global $n=1$
vortices. This allows us to construct arbitrary mixtures of doubly and singly
quantized vortices and to investigate the processes of phase
separation in this two-component Coloumb plasma \cite{Avilov}.

As distinct from the singular to continuous or  continuous to singular
transformations, which was never realized, the transformation of the
dipole-locked continuous to its dipole-unlocked modification has been recorded
\cite{DLtoDUtransformation}. According to our current understanding the
transition between two continuous vortices is mediated by a point-like object -
a hedgehog or  monopole in the ${\bf \hat d}$-field - which represents the
interface between two topologically different pieces of the vortex line
(Fig.~6). Since the vortices on either side of their interface have different
$\pi_2$ topological charges,
$\nu_d=0$ and $\nu_d=1$, the interface between them is the ${\bf \hat
d}$ hedgehog, which is described by the nontrivial element $\nu_d=1$ of the
homotopy $\pi_2(M_A)=Z$. This combination of a linear and a point defect
is the counterpart of a string terminating on a monopole in particle
physics \cite{Vilenkin}. In the NMR experiments, we guess that the
hedgehog is created on the bottom  (or top) wall of the container and moves
up (or down) the vortex line, transforming one vortex into another
(Fig.~7).

In the context of the electroweak model, transitions between vortices
of different winding (topological charge) have been considered in Ref.
\cite{aaetal}.

\section{Fermions and gauge fields} \label{Fermions}

We now come to the second level of analogies as described in the
beginning of Section 2. This is the analogy between the fermionic
sector of the electroweak model and the interactions of quasiparticles
in $^3$He-A with the order parameter.

\subsection{Fermionic sector of the electroweak model}

The fermionic sector Lagrangian
of the standard model of the electroweak interactions
is as follows:
$$
L_f = L_l + L_q
\eqno(4.1.1)
$$
where, the lepton and quark sector Lagrangians for a single family are:
$$
L_l = - i {\bar \Psi} \gamma^\mu D_\mu \Psi
      - i {\bar e}_R \gamma^\mu D_\mu e_R
      + h({\bar e}_R \Phi^{\dag} \Psi + {\bar \Psi} \Phi e_R )\eqno(4.1.2)
$$
$$
L_q = -i ({\bar u} , {\bar d})_L \gamma^\mu D_\mu
         \pmatrix{u\cr d\cr}_L
      -i {\bar u}_R \gamma^\mu D_\mu u_R
      -i {\bar d}_R \gamma^\mu D_\mu d_R
$$
$$
-G_d \biggl [ ({\bar u}, {\bar d})_L \pmatrix{\phi^+\cr \phi\cr} d_R
+{\bar d}_R (\phi^{-} , {\phi^*}) \pmatrix{u\cr d\cr}_L \biggr ]
$$
$$
 - G_u \biggl [ ({\bar u}, {\bar d})_L
   \pmatrix{-{\phi^*}\cr \phi^{-}\cr} u_R +
      {\bar u}_R (-\phi, \phi^+) \pmatrix{u\cr d\cr}_L
       \biggr ]
\eqno(4.1.3)
$$
with $\phi^- = ( \phi^+ )^*$
and
$$
\Psi = \pmatrix{\nu\cr e\cr}_L \ .
$$
The indices $L$ and $R$ refer to left- and right-handed components.

In our analysis we will only be dealing with one fermion family at
a time and hence we shall not be considering the effects of family
mixing such as occurs due to the Kobayashi-Maskawa fermion mass
matrix \cite{chengli}.

The covariant derivatives occuring in the electroweak Lagrangian
are:
$$
D_\mu \Psi = D_\mu \pmatrix{\nu \cr e\cr}_L =
\biggl ( \partial_\mu - {{i } \over 2} \tau^a W_\mu ^a +
{{i } \over 2}  Y_\mu \biggr ) \pmatrix{\nu \cr e\cr}_L
\eqno(4.1.4)
$$
$$
D_\mu e_R = ( \partial_\mu + i  Y_\mu ) e_R
\eqno(4.1.5)
$$
$$
D_\mu \pmatrix{u\cr d\cr}_L =
\biggl ( \partial_\mu - {{i } \over 2} \tau^a W_\mu ^a -
 {{i } \over 6}  Y_\mu \biggr ) \pmatrix{u\cr d\cr}_L
\eqno(4.1.6)
$$
$$
D_\mu u_R = ( \partial_\mu - {{i2 } \over 3} Y_\mu ) u_R
\eqno(4.1.7)
$$
$$
D_\mu d_R = ( \partial_\mu + {{i } \over 3} Y_\mu ) d_R \ .
\eqno(4.1.8)
$$

\subsection{Fermions in $^3$He}

In $^3$He-A, the fermion spectrum differs from that in the electroweak
model. In place of the various quarks and leptons within a single family,
there are only four species occuring as left and right weak doublets.
One way to write these might be
$$
\Psi_L = \pmatrix{\nu\cr e\cr}_L \ ,
\Psi_R = \pmatrix{\nu\cr e\cr}_R \ .
$$
In this section we discuss how this is obtained.

The pair-correlated systems (superconductors and $ ^3$He superfluids) in
their unbroken state above $T_c$ are Fermi liquids: they contain only
interacting fermions. In terms of the field operator $\psi_{\alpha}$
for $^3$He particles the action is
$$
S=\int dt~d^3x~ \left[ \psi^\dagger_{\alpha}
  \left ( i\partial_t - {{{\vec  p}^{~2}} \over {2m}}+\mu \right)
                       \psi_{\alpha} \right ]  ~+~S_{int}
               \ ,
\eqno(4.2.1)
$$
where $S_{int}$ includes the time-independent interaction of two
particles (the quartic term), $m$ is the mass of particle,
$\vec  {\bf p}=-i\vec \nabla$ is  the momentum and $\mu$ is the chemical
potential.  In general this system is described by the large number of
fermionic degrees of freedom. However, as in the bosonic sector of
$^3$He described in Sec. 2.3, there are some simpler limiting cases.

(1) The hydrodynamic case describes the low-frequency motion
of the system. In this case the system of interacting fermions can be
described by only a few degrees of freedom corresponding to the slow
collective motion: particle density $n_N$, spin density
$\vec S$, mass velocity $\vec v$ and entropy density $S$. Their dynamics is
governed by the closed system of hydrodynamic equations.

(2) In the Fermi-gas limit - the limit of small particle-particle
interaction - $^3$He is a simple ensemble of noninteracting particles.
For positive $\mu$, the ground
state is one in which all negative energy fermionic states (with
${\vec p}^{~2} < 2m\mu$) are occupied and corresponds to a {\it solid} Fermi
sphere of radius $p_F = \sqrt{2 m\mu}$.

(3) In the low-temperature limit, the large number of
fermionic degrees of freedom of $^3$He is effectively reduced and
the system is well described as a system of
noninteracting quasiparticles (dressed particles) which, according to
Landau theory, occupies the same Fermi-sphere as a system of noninteracting
particles.
The particle-particle interaction simply renormalizes the effective mass
and the magnetic moment of the quasiparticle. The residual
interaction is reduced at low $T$ because of the small number of thermal
excitations above the Fermi-surface and can be neglected. Thus the action
becomes
$$
S=\int dt~d^3x~\psi^\dagger_{\alpha}[i\partial_t -
\epsilon ( {\vec  p} ) ]\psi_{\alpha}~~,
\eqno(4.2.2)
$$
where $\epsilon ( {\vec  p} )$ is the quasiparticle energy spectrum. In a
Fermi-liquid this description is valid in the so called degeneracy limit,
when the temperature $T$ is much smaller than the effective Fermi
temperature $T_F\sim \hbar^2 /(m a^2)$, where
$a$ is the interparticle distance. This situation takes place almost for all
Fermi-systems above transition since $T_c\ll T_F$. Further for simplicity
we use
$$
\epsilon (\vec  p)=p^2/(2m)-\mu \ .
\eqno (4.2.3)
$$

Below the superfluid/superconducting transition temperature $T_c$, new
collective degrees of freedom appear, which are the order parameter fields,
corresponding to the Higgs field in particle physics. The interaction of
the fermionic degrees of freedom (quasiparticles)  with the order parameter
in the broken symmetry state is described by the action:
$$
S=\int dt~d^3x~\psi^\dagger_{\alpha}[i\partial_t -
\epsilon ( {\vec  p} ) ]\psi_{\alpha}
$$
$$
+{1\over 2}\int dt~d^3x~d^3y~[\psi^\dagger_{\alpha}(x){\bf
\psi}^\dagger_{\beta}(y) \Delta_{\alpha\beta}(x,y) +
\Delta_{\alpha\beta}^\star (x,y)\psi_{\beta}(y)
\psi_{\alpha}(x)]~~.
\eqno(4.2.4)
$$
In the BCS model this action is obtained from Eq.(4.2.1) in the limit when
the particle-particle interaction is small and  only that part
of the interaction is left which leads to the formation of the order
parameter. The quartic interaction in $S_{int}$ is decomposed in a
Hubbard-Stratonovich manner into the bilinear interaction given in
eqn. (4.2.4) where a vacuum expectation value of the product of two
annihilation operators
$$
\Delta \propto <{\rm vac}\vert  \psi \psi
\vert {\rm vac}>~~,
\eqno(4.2.5)
$$
represents the order parameter in the broken symmetry state.
($\Delta$ is a
$2\times 2$ matrix and is not to be confused with the gap amplitude
$\Delta_0$ of Sec. 2.).  $\Delta$ breaks the
$U(1)_N$ symmetry $\psi_{\alpha}\rightarrow e^{i\phi}{\bf
\psi}_{\alpha}$, since under $U(1)_N$ it transforms as $\Delta(x,y)
\rightarrow e^{2i\phi}\Delta(x,y)$. If $\Delta$ has nontrivial spin and
orbital structure, it also breaks the $SO(3)_L$ and $SO(3)_S$
symmetries.

The action in Eq.(4.2.4) allows transition between the states differing by
two particles, say, $N$ and $N-2$. The order parameter  $\Delta$ serves
as the matrix element of such transition. This means that the single-fermion
elementary excitation in the broken symmetry vacuum represents some mixture of
the $N=1$ (particle) and $N=-1$ (hole) states.

In electroweak theory the
interactions corresponding to those in (4.2.4) are the Yukawa interactions
between the left-handed $SU(2)_T$ doublets and the right-handed fermion
singlets, which appear in the broken symmetry state (see Sec. (4.1)).
An example of such an interaction is the term:
$$
G_d ({\bar u}, {\bar d})_L \Phi d_R
$$
in the electroweak Lagrangian.
When $\Phi$ acquires a vacuum expectation value during the electroweak
phase transition, this gives rise to the nonconservation of the isospin
and hypercharge in the same manner as the charge $N$ is not conserved in
the broken symmetry action (4.2.4). These  terms also lead to the lepton
and quark masses. In superfluids and superconductors these terms give
rise to the gap in the quasiparticle spectrum.

In the electroweak vacuum one has many matrix elements
$G_d  \Phi$,  $G_e \Phi$, $G_\mu  \Phi$ , $G_\tau  \Phi$ etc.
which give rise to masses of all quarks and leptons (except for the
left-handed neutrino). All of them have the same symmetry structure.
In condensed matter such a situation - with many symmetrically equivalent
matrix elements in the broken symmetry state - is reminiscent of the
situation in metals with several electron bands,
$\epsilon_a(\vec  p)$, where $a$ is the band index. In such metals, after the
transition into the superconducting state, the matrix elements
$\Delta_{aa'}$ appear between electrons and holes from different bands.
All the elements are characterized by the same broken symmetry.
However in most cases one can distinguish the matrix
element $\Delta_{a_0a_0}$, actually the largest one,  which is ``primary''.
This means that the symmetry breaking Cooper pairing occurs first between the
electrons within the band $a=a_0$. This symmetry breaking induces the other
(smaller) matrix elements $\Delta_{a_0a}$ and $\Delta_{aa'}$ with
$a,a'\neq a_0$.

In some technicolour theories of the dynamical electroweak transition it
was suggested that  the ``primary'' matrix element is also the largest
one, i.e.  the symmetry  breaking  results from the ``Cooper'' pairing
of the heaviest quarks and antiquarks (top quark condensate
\cite{technicolour}), while the
other (smaller) elements are induced by this symmetry breaking.

For the  $s$-wave spin-singlet pairing in superconductors and
the  $p$-wave spin-triplet pairing in superfluid $^3$He the matrix element
has the following general form:
$$
\Delta_{ s-{\rm wave}} (\vec  r,\vec  p)=  i \sigma ^{(2)} \Psi(\vec  r)
{}~~,
\eqno(4.2.6)
$$
$$
\Delta_{ p-{\rm wave}} (\vec  r,\vec  p)=  i \sigma ^{(\mu )}\sigma ^{(2)}
A_{\mu \,i}(\vec  r)   p_i/p_F~~.
\eqno(4.2.7)
$$
Here $\vec \sigma$ are the Pauli matrices in spin space,
$\vec  r=(\vec  x+\vec  y)/2$ is the center of mass coordinate of the
Cooper pair, while the momentum $\vec  p$
describes the relative motion of the two fermions
within the Cooper pair: it is the Fourier transform of  the coordinates
$(\vec  x-\vec  y)$. The complex scalar function $\Psi(\vec  r)$ and
$3\times 3$ matrix $ A_{\mu \,i}(\vec  r)$ are the order parameters for the
two systems.
The symmetries of $^3$He and the A-phase and the order parameter have been
described in Sec. (2.1).

The easiest way to treat the action in Eq.(4.2.4), in which the states with 1
particle and -1 particle are hybridized by the order parameter, is to double
the number of degrees of freedom introducing the antiparticle  (hole) for each
particle.  Let us introduce  the  Bogoliubov-Nambu field operator
$\chi$ which is the spinor in this new  particle-hole space:
$$
{\bf  \chi}=\pmatrix{{\bf u}  \cr
                {\bf v} \cr}=\pmatrix{\Phi  \cr
                i\sigma ^{(2)}\Phi^\dagger\cr}~~;~~
{\bf  \chi}^\dagger =({\bf u}^\dagger~, ~ {\bf v}^\dagger )~~.
\eqno(4.2.8)
$$
It transforms under $U(1)_N$ symmetry  operation
$\Phi_{\alpha}\rightarrow e^{i\phi}\Phi_{\alpha}$ as
$$
{\bf  \chi}\rightarrow
e^{i\check \tau _3 \phi} {\bf  \chi}~~,
\eqno(4.2.9)
$$
where $\check \tau _i$ are Pauli matrices in the Nambu space,
such that $\check \tau_3$ is the operator ${\bf N}$ for quasiparticles
with the eigenvalue $+1$ for the particle and $-1$ for the hole.

The eigenvalue equation for the quasiparticle spectrum is
$$
{\cal H}  \chi =E  \chi~~ ,
\eqno(4.2.10)
$$
with the Hamiltonian
$$
{\cal H}_{ s-{\rm wave}}= \epsilon(\vec  {\bf p})\check \tau
_3 + \pmatrix{0 &\Psi(\vec  r) \cr
              \Psi^*( \vec  r) & 0 \cr }~~.
\eqno(4.2.11)
$$
in $s$-wave superconductors, and,
$$
{\cal H}_{\rm A-phase}= \epsilon(\vec  {\bf p})\check \tau
_3 + {{\Delta _0}\over {p_F}}(\vec   \sigma\cdot\hat{\bf d})(\check \tau
_1\hat e_1(\vec  r)\cdot \vec  {\bf p} -\check \tau
_2\hat e_2(\vec  r)\cdot \vec  {\bf p}  ) ~ ,
\eqno(4.2.12)
$$
in the A-phase of $^3$He.

The square of the fermion energy
$$
E_{ s-{\rm wave}}^2(\vec  p)=
{\cal H}^2=\epsilon^2(\vec  p)+\vert \Psi\vert^2~~ ,
\eqno(4.2.13)
$$
$$
E_{\rm A-phase}^2(\vec  p)={\cal H}^2=\epsilon^2(\vec  p)+({{\Delta _0}\over
{p_F}})^2(\vec  p \times {\bf \hat  l})^2~~ ,
\eqno(4.2.14)
$$
together with (4.2.3),
shows that the fermions in the $s$-wave system have a gap in the spectrum,
while in the A-phase, if the chemical potential $\mu$ is positive,
the quasiparticle spectrum has two zeros (nodes): at
$\vec  p=\pm p_F\hat{\bf l}$. These nodes are the source of anomalies in
$^3$He-A.

In the vicinity of each zero the spectrum of the fermions
is relativistic and from  Eq.(4.2.12) it follows
$$
{\cal H}=\sum_{N=1}^3  e^j_N {\bf{\check \tau}}_N (p_j-q_A A_j) \ ,
\eqno(4.2.15)
$$
where $\vec  A=p_F\hat{\bf l}$ plays the role of a vector potential like an
``electromagnetic'' field, $q_A$ is the corresponding ``electric''
charge which is $+1$ for the fermions in the vicinity of the node
$\vec  p=p_F\hat{\bf l}$ and $-1$ for the
fermions in the vicinity of the opposite node at $\vec  p=-p_F\hat{\bf l}$.
The coefficients $e^i_N$   are
$$
\vec  e_1=2S_3{{\Delta_0}\over{p_F}}\hat e_1 \ , \
\vec  e_2=-2S_3{{\Delta_0}\over{p_F}}\hat e_2 \ , \
\vec  e_3=q_Av_F\hat{\bf l}  \ \  ,
\eqno(4.2.16)
$$
where  $v_F = p_F /m$ is the Fermi velocity; the quantum number $S_3=\pm 1/2$
is the spin projection of the fermions on the axis $\hat{\bf d}$.

The coefficients $e^i_N$ form the so called $dreibein$,
or triad, the local coordinate frame for the
fermionic particles. They are the 3-dimensional version of the
$vierbein$ or tetrads, which are used to describe gravity
in the tetrad formalism of general relativity. From Eq.(4.2.16) it follows
that $det[ e^i_N ]$ has the sign of $q_A$,
{\it i.e.} the $A-$charge, $q_A$, also defines the parity of
the fermions: the fermions with positive $q_A$ are right-handed and
those with negative $q_A$ are left-handed.

The conventional metric tensor expressed in terms of the
triad $e^i_N$ is
$$
g^{ij} =\sum_{N=1}^3   e^j_N e^i_N \ ,
\eqno(4.2.17)
$$
so the square of the energy of fermions in the vicinity of each of the
nodes is
$$
E^2(\vec  p)=g^{ij}(p_i-q_A A_i)(p_j- q_A A_j)  .
\eqno(4.2.18)
$$

Eq.(4.2.15) is the Weyl Hamiltonian for charged chiral particles:
the positively charged left handed fermions are concentrated
near the momentum $\vec  p=p_F\hat{\bf l}$, while the negatively charged
right handed fermions are concentrated near the momentum
$\vec  p=-p_F\hat{\bf l}$. Note that the metric is flat
if $\hat{\bf l}$ is constant.

Note also the close analogy of eqn. (4.2.15) with the interaction
of the electroweak $Z-$field with the fermions. For example,
if we set all other gauge fields ($A$, $W_1$ and $W_2$) to zero,
the interaction
of the left- and right-handed electrons with the $Z$ field is
given by
$$
-i {\bar e}_L \gamma^\mu \left ( \partial_\mu + {i \over 2}
cos(2\theta_w) Z_\mu \right ) e_L
-i {\bar e}_R \gamma^\mu \left ( \partial_\mu + {i \over 2}
(cos(2\theta_w)-1) Z_\mu \right ) e_R \ .
$$
For $sin^2\theta_w = 0.25$ -- which is quite close to the observed
value of 0.23 -- even the $Z$ charges of the left- and right-handed
electrons are related by a minus sign.

\subsection{Bosonic sector of $^3$He-A on Level 2 of analogy}

Inspite of the similarity between $\vec A = p_F \hat {\bf l}$
in Eq. (4.2.15) and the $Z$ gauge field,
$\vec A$ is closer to the {\it massless} electromagnetic field $A$,
than to the {\it massive} $Z-$field.  The $\hat
{\bf l}$-vector is a dynamical field: the propagating oscillations of
$\hat{\bf l}(\vec r,t)$ (the so called orbital waves) have the massless
spectrum of the photons\cite{Exotic}.

It should be mentioned that  the ``electromagnetic'' field $\vec  A$
and ``electric'' charge $e$ on this Level 2 of analogies
has nothing to do with the ``electric'' charge of the analogy at Level 1.
On Level 1 the ``electromagnetic'' field is absent, since the $U(1)_Q$
group is global. On Level 2 the ``gauge'' field
$\vec  A=p_F\hat{\bf l}$ is a part of the order parameter
whose action on the fermions is like the action of the
electromagnetic field on the fermions
in the electroweak model. Actually
$\vec  A=p_F\vec  l$ corresponds to an axial field since it acts in
opposite ways on left-handed and right-handed fermions.

On Level 1 of the analogy, the $W$-bosons should be related to the
$SO(3)_L$ group, but they are absent because this group is global.
However, on Level 2 the electroweak $W$ bosons arise \cite{Exotic}.
The 2 additional gauge fields $W^1$ and $W^2$, corresponding to the
$W$-bosons which are transverse  to the $\hat {\bf d}$
(i.e.  $\hat {\bf d}\cdot \vec {\bf W}=0)$) enter the Hamiltonian for
fermions, if one considers the effect of the spin ${\check  \sigma}$
of the $^3$He atoms
and of the other (non-Goldstone) degrees of the order parameter
$$
A_{\mu \,i}=({\rm A-phase~vacuum})+\delta A_{\mu \,i}~~.
\eqno(4.3.1)
$$
As a result the Hamiltonian for fermions is
$$
{\bf H}=\sum_{N=1}^3~ e^i_N{\bf{\check \tau}}_N (p_i - q_A A_i-q_W\tilde
{\check  \sigma}_{\alpha}
W^{\alpha}_i)\hskip2mm   .
\eqno(4.3.2)
$$
Here $q_W=q_A$; the components of the $ W^{\alpha}_i$ field are related to
$\delta  A_{\alpha i}=u_{\alpha i}+iv_{\alpha i}$ in the following way:
$$
W^{\alpha}_x + i W^{\alpha}_y= i(p_F/\Delta_A)e^{\alpha \beta\gamma}
\hat d^{\beta}
\delta A_{\gamma z}~~.
\eqno(4.3.3)
$$

So only 4 components of $ W^{\alpha}_i$  are nonzero:
$$
W^{1}_x=v_{yz}p_F ~,~W^{2}_x=-v_{xz}p_F ~,~W^{1}_y=-u_{yz}p_F
{}~,~W^{2}_y=u_{xz}p_F ~~,
\eqno(4.3.4)
$$
which corresponds to the gauge field produced by the transverse
$W$-bosons. The  contribution to the longitudinal $W^3$-bosons is absent
in the $p$-wave order parameter $ A_{\alpha i}$.

There are altogether 18 real elements of the $3\times 3$ matrix $ A_{\alpha
i}$: 4 of them correspond to $W$-bosons; 2 - to electromagnetic field
(photons), 3 to gravitons,
while the other 9 have no analogy in electroweak interactions.

The electromagnetic and the $W$-boson sector of $^3$He-A in this Level 2
analogy  has the form
$$
L_W +L_A = m_W^2 W_{\mu  a} W^{\mu  a}- {1 \over 4 g^2} W_{\mu \nu a} W^{\mu
\nu a} - {1 \over 4 e^2} A_{
\mu \nu} A^{\mu \nu}~~,
\eqno(4.3.5)
$$
where $m_W$ is the mass of the $W$ boson and $g=e$ which corresponds to the
weak mixing angle $\theta_W=\pi/2$.
Both charges $e$ and $g$ experience the zero-charge effect due to
the screening polarization of the fermion vacuum: as $T\rightarrow 0$ one
has $e^2=g^2\propto 1/\ln {\Delta_0\over T}$ which results in the
logarithmic divergence of the gradient energy in Eq.(2.3.5).
This is different from the standard model, where the charge
$g$ experiences the asymptotic freedom. This is because in $^3$He-A the
charged $W$-bosons do not contribute to the renormalization of $g$.

Note that the metric $g_{ij}$ appearing in (4.2.17) appears to
be related to the electromagnetic potential $A_i$ in (4.2.15).
But this relationship in only valid on the vacuum manifold.
If we consider small oscillations of all 18 components of the
order parameter, i.e. including the non-Goldstone modes,
we find that the ``gravitational'' waves and the ``electromagnetic''
waves are independent, at least for propagation along $\hat {\bf l}$.
In this case, the gravitons are (massive) oscillations of
$g_{11} - g_{22}$ and $g_{12}$ and are unrelated to the (massless)
oscillations of $\hat {\bf l}$. So while there are similarities
on this {\bf Level 2} of analogies, there are some differences
too which one must bear in mind.

\section{Anomalies}

\subsection{Baryon number anomaly in the electroweak model}

In the electroweak model described in Section (4.1) there are
additional accidental global symmetries that are present.
These are the $U(1)_B$ and $U(1)_L$ symmetries whose classically
conserved charges are baryon number and lepton number.
Each of the quarks is assigned 1/3 baryon number and zero lepton
number while the leptons (neutrino and electron in Section (4.1))
are assigned lepton number 1 and zero baryon number.

The conservation of baryon number ($B$) and lepton number ($L$) is
anomalous and the current conservation equations are:
$$
\partial_\mu j_B ^\mu = \partial_\mu j_L ^\mu
= {{N_F} \over {32\ \pi^2}} \left (
- W_{\mu \nu}^a {\tilde W}^{a\mu \nu} +   Y_{\mu \nu}
{\tilde Y}^{\mu \nu} \right )
\eqno(5.1.1)
$$
where, $N_F$ is the number of families and the dual of the field strength is
defined by ${\tilde F}^{\mu \nu} = {1 \over 2}
\epsilon^{\mu \nu \lambda \sigma}F_{\mu \nu}$.  Note that
$B-L$ is still conserved but $B+L$ is not. The nonconservation results
from the spectral flow of the chiral massless fermions in the presence of the
field strength because the left-handed and right-handed fermions in
Eqs.(4.1.4-8) have different charges.

In the case of the electroweak strings the spectral flow is provided by
the fermion zero modes localized on strings.
If we consider a configuration of electroweak $Z-$string loops, the direct
integration of the macroscopic anomaly equation (5.1.1) gives
the following change in baryon number between two string configurations
\cite{tvgf}:
$$
\Delta B = 2 N_F cos(2\theta_w )\Delta L \ .
\eqno (5.1.2)
$$
where, $\Delta L$ is the change in the linking number of the
strings. This result can be derived in a more rigorous microscopic
theory by considering the spectrum of fermions in a string background and by
working out the quantum numbers of the ground state of the fermions. This
calculation was done in Ref. \cite{jgtv} by properly regularizing the
divergent terms. However the  calculation for the similar phenomenon in
$^3$He (see Sec. (6.7) below) suggests that the macroscopic result is
true only in the limit of large $B$.

The origin of the non-trivial baryon number of linked loops of string is
due to an Aharanov-Bohm interaction of the fermionic zero modes on one loop
with the other linked loop. This interaction shifts the energy levels of the
fermions and leads to non-trivial quantum numbers.

Below we consider the similar effects
in the $^3$He-A texture  (Sec.5.2), where the macroscopic anomaly equation
of the type of Eq.(5.1.1) is used, and within the core of vortices (Sec.6),
where the microscopic analysis of the fermionic zero modes is presented.

\subsection{Mass current anomaly  in $^3$He-A}

The axial anomaly is a consequence of the spectral flow of fermions:
if the fermions are gapless, the inhomogeneous vacuum induces a spectral
flow, which carries particles from negative energy states of the vacuum
to positive energy states.
As a result, the corresponding conserved quantity (viz, charge)
is  transferred from a coherent vacuum motion into incoherent fermionic
degrees of freedom, which is visualized as creation of charge from the
vacuum. It was
shown in \cite{A-phase anomaly,Stone} that the same
phenomenon takes place in superfluid $^3$He-A, where the gap in the
quasiparticle spectrum has point
nodes. This results in a spectral flow of fermionic quasiparticles
through the nodes when the condensate evolves in time. Since the
left-handed and right-handed fermions have opposite linear momentum an
immediate result of this effect is the linear momentum (or mass current)
anomaly in $^3$He-A.

The mass density $\rho$ in superfluid $^3$He is equal to the bare mass
$m_3$ of $^3$He atom times the particle density
$$
\rho(\vec  r,t)=m_3<{\rm vac}\vert
\Phi^\dagger_{\alpha}\Phi_{\alpha}
\vert{\rm vac}>\  , ~~\int d^3r~\rho(\vec  r,t)=
m_3<{\rm vac}\vert  {\bf N} \vert{\rm vac}> \ .
\eqno(5.2.1)
$$
Using Eq.(4.2.8) the particle density can be also represented in
terms of the Bogoliubov-Nambu field $\chi$ and the charge $N$
operator $\check\tau_3$:
$$
\Phi^\dagger_{\alpha}\Phi_{\alpha}=
              {1\over 2}{\bf \chi}^\dagger \check \tau_3 {\bf \chi}~~,
\eqno(5.2.2)
$$
the factor $1/2$ compensates for the double counting in the description
in terms of particles and holes. If all the fermionic excitations of
superfluid vacuum are massive (i.e. there are no nodes), the slow
dynamics of the superfluids or superconductors at
zero temperature is the dynamics of the
collective variable (the order parameter field and the density $\rho$) since
the fermionic quasiparticles are not created. The mass (particle) conservation
law (which comes from the $U(1)_N$ symmetry of physical laws) is thus
exhausted by the collective  (hydrodynamic) variables:
$$
\partial_t\rho +\vec \nabla\cdot\vec  j =0~~,
\eqno(5.2.3)
$$
where the mass current (or the density of linear momentum) in the
nodes-free superfluids is
$$
\vec  j_{\rm node-free}  =
\rho\vec  v_s+{1\over 2}\vec \nabla\times \vec {\bf L} ~~.
\eqno(5.2.4)
$$
Here $\vec  v_s$ is the  superfluid velocity. The first term in
(5.2.4) is the zero temperature limit of (2.1.8a): the density
$\rho_n$ of the thermal fermionic excitations is zero at $T=0$. The
second term is the orbital current originating from the internal angular
momentum of the liquid. The vector $\hat{\bf L}$ is the density of the
angular momentum. For the hypothetical nodes-free A-phase state
(i.e. for $\mu<0$) the symmetry  ${\bf Q}={\bf L}_3-(1/2){\bf N}\equiv
{\bf L}\cdot\hat{\bf l}-(1/2){\bf N}$ of the vacuum requires that the
angular momentum density equals $1/2$ of particle density,
since the charge $Q$ of the vacuum is zero:
$$
\vec{\bf L}_{\rm node-free}  ={\hbar\over 2m_3}\rho~\hat{\bf l}~~.
\eqno(5.2.5)
$$

The gap nodes, via the axial anomaly, modify the  equation (5.2.4) for the
mass current.
Also the vacuum can accumulate the charge  $Q$
of gapless fermions, which leads to the nonzero charge $Q$ of the
vacuum. As a result the  angular momentum can deviate from its
fundamental form in Eq.(5.2.5).

The gapless chiral  fermions in Eq.(4.2.15) obey the same
Adler-Bell-Jackiw anomaly equation \cite{Adler,BellJackiw} which gave rise
to Eq.(5.1.1) for the baryonic and leptonic charges. Here we are interested
only in the effect produced by the ``gauge'' field
$\vec  A=p_F\hat{\bf l}$:
$$
\partial_t   Q_c=
           {1\over {2\pi^2}}\int d^3r ~(\partial_t \vec  A \cdot
                  (\vec  \nabla \times \vec A  \, \, ))  ~~,
\eqno(5.2.6)
$$
where $Q_c=\int d^3r~ \delta S/\delta A_0$ is the chiral charge of the
fermions. This anomaly equation describes the production of the chiral
charge in $^3$He-A during the dynamics of the order parameter $\vec
A=p_F\hat{\bf l}$.

Since each created left fermion carries the linear momentum
$p_F\hat{\bf l}$ and the same momentum is carried by the created left
antifermion, the total production of the quasiparticle linear momentum
is\cite{A-phase anomaly}:
$$
\partial_t  \vec  P_{qp}=
     {1\over {2\pi^2}}\int d^3r~ p_F\hat{\bf l}~(\partial_t \vec  A
            \cdot (\vec  \nabla \times \vec  A  \, \, ))  ~~.
\eqno(5.2.7)
$$
This is a linear momentum anomaly in $^3$He-A.

Since the total linear
momentum is nevertheless conserved, Eq.(5.2.7) means that momentum is
transferred from the collective variables describing the inhomogeneous
vacuum to the system of quasiparticles.

This equation allows us to calculate the extra mass current of the
$\hat{\bf l}$ texture due to accumulation of the fermionic charge by the
texture, which would correspond to the baryonic charge in
Eq.(5.1.2) accumulated by the inhomogeneous vacuum of $Z$-string.
Let us take
the arbitrary but fixed $\hat{\bf l}(\vec  r)$ texture and consider the
process in which the node-free A-phase transforms to the ``real'' A-phase,
that is, the phase with nodes. This  occurs if the Fermi momentum changes from
$p_F=0$ at $t=t_0$ to its equilibrium value in the real A-phase. In this
process
$\partial_t \vec  A = \hat{\bf l}\partial_t p_F$. The
total momentum of the texture thus changes according to Eq.(5.2.7):
$$
\vec  P(t)-\vec  P(t_0) = -\int_{t_0}^t dt~\partial_t  \vec  P_{qp}
  = - {1\over {6\pi^2}} \int d^3r~ p_F^3 ~\hat{\bf l}~
               (  \hat{\bf l} \cdot (\vec  \nabla \times \hat{\bf l} )) ~,
\eqno(5.2.8)
$$
where $\vec  P(t_0)$ is the anomaly-free momentum
$$
\vec  P(t_0)=\int d^3r~ \vec  j_{\rm node-free}~~.
\eqno(5.2.9)
$$
The mass current in the A-phase (the momentum density) thus becomes
$$
\vec j =\vec  j_{\rm node-free}+\vec  j_{\rm anomalous} =
\rho \vec  v_s  +
{1\over 2} \vec  \nabla \times {{\rho\hat{\bf l}}\over {2m_3}}
- {\hbar\over 2m_3} C_0  \hat{\bf l} \bigl( \hat{\bf l} \cdot
(\vec  \nabla \times \hat{\bf l} ) \bigr)
\eqno(5.2.10)
$$
where, $C_0=m_3p_F^3/3\pi^2$.
The extra mass current results from the helicity of the $\vec  A$ field:
$$
\vec  j_{\rm anomalous} =  -{1\over {6\pi^2}}  \,   p_F\hat{\bf l} \,
\bigl( \vec A \cdot (\vec  \nabla \times \vec A \, \, ) \bigr)  ~~,
\eqno(5.2.11)
$$
in the same manner as the baryonic charge of string is the consequence of
the helicity of the gauge fields \cite{tvgf}.

\subsection{Momentum exchange between the  moving continuous vortex and
heat bath. }

The axial anomaly results in a curious exchange of the linear momentum
between the moving texture and the heat bath. Let us consider this on the
example of the $n=2$ continuous vortex discussed in Sec.(3.3). When this
vortex moves with velocity $\vec v_L$, the $\hat{\bf l}$-texture becomes
time dependent: $\hat{\bf l}=\hat{\bf l}(\vec r-\vec v_Lt)$.
As a result the ``electric'' field arises
$$
\vec E=\partial_t \vec A=p_F\partial_t \hat{\bf l}=-p_F(\vec
v_L\cdot\vec\nabla)
\hat{\bf l} \, \,  ,
\eqno(5.3.1)
$$
According to the anomaly equation (5.2.7) this leads to the production of
quasiparticle momentum. This momentum is absorbed by the normal component of
the liquid (the heat bath) moving with the velocity $\vec v_n$. This means
that there appears a (reactive, i.e. nondissipative) force  between the vortex
and the heat bath caused by the spectral flow of the fermions. The force
acting per unit length of the vortex is
$$
\vec F_{\rm spectral~flow} =
     {1\over {2\pi^2}}\int dx~dy~ p_F^3\hat{\bf l}~(((\vec
v_L-\vec v_n)\cdot\vec\nabla)
\hat{\bf l}
            \cdot (\vec  \nabla \times \hat{\bf l}  \, \, ))  ~~,
\eqno(5.3.2)
$$
Here $\vec v_n$ enters since for this effect the heat bath reference frame
is the relevant frame for the vortex motion.

Simple transformation of this equation using the integration by parts gives
\cite{HydrodynamicAction}
$$
\vec F_{\rm spectral
{}~flow}={C_0\over 2m_3} (\vec v_L-\vec v_n)\times \hat z \int dx~dy~\hat{\bf
l}\cdot (\partial_x \hat{\bf l}
\times \partial_y\hat{\bf l})    ~~.
\eqno(5.3.3)
$$
Finally using the  Eq.(3.3.2) one obtains the anomaly contribution to the
force acting on the $n=2$ vortex:
$$
\vec F_{\rm spectral
{}~flow}=2 C_0\vec \kappa\times (\vec v_n-\vec v_L) ~~.
\eqno(5.3.4)
$$

This force does not depend on the details of the vortex structure and is
defined by the anomaly parameter $C_0$ in Eq.(5.2.10) and by the winding
number $n$ of the vortex. This stresses the topological origin of this
anomalous force.

In the following Section it will be shown that  the same type of
anomaly (and the same force) arises for vortices in any Fermi superfluid  or
superconductor, even if they do not contain the gap nodes in homogeneous
state. For conventional singular vortex, say, in $s$-wave superconductor, the
effect occurs because the quantized vortex itself, due to its  singularity at
the vortex axis, plays the same role as gap nodes in  A-phase of $^3$He
\cite{Q-modes-Index}. The microscopic analysis shows that the vortex
gives rise to the fermion zero modes in the vortex core, whose spectral flow
is responsible for the anomalous force. The result of this microscopic
calculation agrees with Eq.(5.3.4) obtained from the  macroscopic anomaly
equation, but only in a special limit.

\section{Fermions on vortices and strings}\label{Fermions-on-strings}

\subsection{Symmetry generators} \label{SymmetryGenerator}

Let us consider singular vortices in the $s$-wave superconductors and
in the A-phase, and singular $Z$-strings in the electroweak vacuum
(Eq.(2.3.7)). Here we are interested in the simplest vortices, which
display the cylindrical symmetry. The asymptotical behavior of the order
parameter far from
the vortex core of  coherence length size ($\xi(T)\sim 1/m_{Higgs}(T)$ in
electroweak theory) is
$$
\Phi_{\rm ew}(r \rightarrow \infty) = {\eta \over {\sqrt{2}}}  e^{in\phi}
           \left(\matrix{ 0\cr 1\cr}\right)~~,
\eqno(6.1.1a)
$$
$$
\vec  \Psi_{\rm {}^3He-A}(r \rightarrow \infty)= e^{ in\phi}~
          {{\hat x+i\hat y}\over \sqrt 2}~~,
\eqno(6.1.1b)
$$
$$
\Psi_{s-{\rm wave}}(r \rightarrow \infty)=\Delta_0 e^{in\phi} ~~.
\eqno(6.1.1c)
$$
All these linear defects   have an integer winding number $n$, except for
the  Alice string in the A-phase, which has $n=\pm 1/2$. In the A-phase and
electroweak vacuum these linear defects have a constant $\hat{\bf l}=\vec z$;
the vortices and strings, which correspond to each other, have the following
relation between their winding numbers: $n_{\rm Z-string}=(1/2) n_{ \rm
vortex}$.
This means that a $4\pi$ $^3$He-A vortex corresponds to a $2\pi$ $Z-$string
(both are topologically unstable and can transform to the disgyration in the
$\hat{\bf l}$ field with winding number $n=2$). For the $2\pi$-vortex  and
for the
$\pi$-vortex (the Alice string) there are no counterparts in the electroweak
vacuum. The Alice string exists in the A-phase, only if one takes into account
the total order parameter including the spin degrees of freedom (see
eqn.(3.1.5b)). The asymptotic form of the order parameter  for the 2$\pi$
vortex is given in eqn. (3.1.5a).

The electroweak $Z-$string in Eq.(6.1.1a), in addition to the
translational symmetry along the vortex axis with the generator ${\bf \vec
p}_z$, has two continuous symmetries  given by the following generators:
$$
{\bf Q}={\bf T}_3+(1/2){\bf Y}~~, ~~{\bf Q}_n={\bf L}_3-n {\bf Y}~~;
\eqno(6.1.2a)
$$
where ${\bf L}_3=-i\partial_\phi$ is the generator of the coordinated
rotations.

For vortices in conventional superconductors one also has two
generators:
$$
{\bf Q}_n={\bf L}_3-{{n }\over 2}{\bf N}~~,~~
{\bf S}_3 ~~;
\eqno(6.1.2b)
$$
Here  ${\bf S}_3$ is the generator of the  $SO(3)_S$ rotation group, which  is
not broken in $s$-wave superconductors.

Two generators also take place in the axisymmetric A-phase vortices with
integer winding number $n$:
$$
{\bf Q}_n={\bf L}_3-{{n +\hat l_z(\infty)}\over 2}{\bf N}~~,~~
{\bf S}_3 ~~,~~n={\rm integer}~~;
\eqno(6.1.2c)
$$
Here $\hat l_z(\infty)=\pm 1$ gives the orientation of $\hat{\bf l}$
vector at infinity;
${\bf S}_3$ is the generator of the  $SO(2)_S$  rotations
about the vector $\hat{\bf d}$, i.e. of the little group of the
broken symmetry
group $SO(3)_S$ in Eq.(2.1.1b), and,  ${\bf L}_3$ includes both the
coordinate rotations $ -i\partial_\phi$ and internal rotations of the orbital
part of the order parameter $\vec \Psi$ (see Eq.(3.1.8)).
(Rembember that in $^3$He, orbital rotations in $SO(3)_L$ are not just
rotations in ``internal'' space as in the electroweak model; instead
these are rotations in physical space as well (see eqn. (3.1.8c)). Then,
axial symmetry implies that the $z-$component of $\hat {\bf l}$
is $\pm 1$ at infinity.)

For the Alice strings in
$^3$He-A, {\it i.e.} vortices with  half integer winding
number, one has only one generator, since the orbital and spin degrees of
freedom are coupled topologically:
$$
{\bf Q}_{n}={\bf L}_3+
       {1\over 2}{\bf S}_3-{{n+\hat l_z(\infty)}\over 2}{\bf N}~~,~~
         n=\pm 1/2~~.
\eqno(6.1.2d)
$$

These symmetries, if they are  not spontaneously broken in the vortex core,
also define the vortex structure in the core. For example, Eq.(6.1.2b) implies
that the order parameter for the symmetric vortex in the $s$-wave
superconductor is everywhere described by one radial function
$$
\Psi_{s-{\rm wave}}(\vec r)\equiv \Delta(\vec r)=\vert \Delta(r)
\vert e^{in\phi} ~~,
\eqno(6.1.3)
$$
while the symmetric A-phase vortex in Eq.(3.1.5c) contains 2 radial functions.

For typical vortices realized and observed in superfluid $^3$He phases some
of the vortex symmetries are spontaneously broken, including the continuous
symmetries. The breaking of axial symmetry in the vortex core was
experimentally observed in $^3$He-B vortices due to the new Goldstone mode
\cite{CoreAsymmetry}.  In cosmic strings the analogous spontaneous breaking of
the  continuous symmetry in the vortex core has been discussed by Witten
\cite{Witten}. In this case the spontaneously broken symmetry is the
electromagnetic symmetry $U(1)_Q$ with the generator ${\bf Q}$ in Eq.(6.1.2a),
which implies the appearance of superconductivity within the core of the
string with a nondissipative current along the vortex axis.

\subsection{Fermion Zero Modes and Symmetry of the Vortex}

The existence of low-energy fermions (``zero modes'') in a vortex
background can be deduced by the application of certain index theorems
which relate the number of such modes to the vortex winding number
$n$. Originally this relation was found for the  fermionic spectrum localized
on strings in particle physics \cite{Jackiw,Hindmarsh}.

The spectrum of single-fermionic excitations in a vortex,
$$
E_{n_r}(p_z, Q ,Q_n, S_3,..)
$$
depends on the momentum projection $p_z$ on the vortex
axis (a continuous quantum number), and on the discrete eigenvalues
of the generators of continuous symmetry: these quantum numbers can
be integer or half integer. $n_r$ here is the radial quantum number,
and is not due to any symmetry of the system.

In relativistic theories the number of branches $E(p_z)$ crossing
$E=0$ as functions of $p_z$ equals $n$ (see Fig.~8a
for $n=1$). For fermions in condensed matter vortices there is no such
theorem.
However a similar theorem exists if one considers the spectrum as a
function of the generalized angular momentum $Q_n$ in Eqs.(6.1.2b-d)
Fig.~8b-d. The interlevel distance of bound states
$\partial E_{Q_n} /\partial Q_n=\omega _0$
is small compared to the gap amplitude $\Delta_0(T)$ of fermions in bulk:
$\omega_0\sim \Delta_0^2(T)/E_F\ll \Delta_0(T)$. Thus, if one neglects the
interlevel distance as compared, say, with temperature or with  the level
width, the spectrum can be considered as a continuous function of the
{\it continuous } parameter $Q_n$. As a function of continuous $Q_n$ the
spectrum has anomalous  (chiral) branches, fermion zero modes, $E_0(p_z,Q_n)$
whose number $N_{\rm zm}$ is related to the vortex winding number
$N_{\rm zm}=2N$ according to the index theorem \cite{Q-modes-Index}.
As a function of $Q_n$, each anomalous branch crosses zero of energy an
odd number of times and runs through both discrete and contunuous spectrum
from $E_0=-\infty $ to $E_0=+\infty $. Any other branch either does not
cross zero of energy at all or crosses it an even number of times.
For low-energy bound states, the spectrum of the chiral branch is linear
in $Q_n$.  For the most symmetric  vortices, for example,
$$
E_0(p_z,Q_n)=Q_n\omega_0(p_z) ~~.
\eqno(6.2.1)
$$

In the next subsection this will be derived in a simplified way with the aid of
the quasiclassical approximation valid in the limit $Q\gg 1$, ie in the range
where we can consider $Q_n$ as continuous variable. Thus,
in the limit of large $Q_n$, when $Q_n$ can be considered as a continuous
coordinate in the 2D  momentum space $(Q_n,p_z)$, the fermions occupying the
negative energy levels form a 2D Fermi  liquid. The role of the Fermi surface
is
played by the line $Q_n=0$, at which the energy spectrum crosses zero.

However, as was found in original paper \cite{Caroli} for $s$-wave
superconductors, the Eq.(6.2.1) is valid even for small $Q_n$ , where the
discrete nature of $Q_n$ becomes important. If one is interested in the fine
energy scale of the order or less than the interlevel distance $\omega_0$, one
again comes  to the problem of existence of the gap nodes in the spectrum as
function of $p_z$. Three different types (a-c) of the  behavior of the
fermionic
spectrum in the low-energy  limit are posssible (Fig.8~b-c) \cite{MisVol}.
One of
the factors which determines the behavior of the spectrum at low energy is the
following property of the quantum number $Q_n$ of the fermions:
$Q_n$ can be either half-integer or integer depending on the vortex symmetry.
It appears that for    conventional $n=1$ vortices in $s$-wave superconductors
the quantum number $Q_n$ is half-integer. Then, according to the Eq.(6.2.1) the
case (a) occurs: there is a  finite gap  $E(Q=1/2,p_z=0)= \omega_0/2$ in the
fermionic spectrum (Fig.~8b).

In the most symmetric $^3$He-A vortices the orbital momentum $Q_n$ is integer.
This leads to a possibility of the case (b): the branch with $Q_n=0$
represents the flat band with zero energy: $E(Q_n=0,p_z)\equiv 0$ for all $p_z$
(Fig.8d). Existence of such flat band is supported by the symmetry of the
vortex  and is also confirmed in exact calculations \cite{KopSal} of the
fermion
spectrum.

In the most symmetric $^3$He-B vortex the symmetry, which leads to zero
energy at $Q_n=0$, takes place only at $p_z=0$. This produces the case (c):
One
or more branches of the spectrum with particular $Q_n$'s (in our case with
$Q_n=0$) cross the zero level as functions of
$p_z$ (Fig.8c). In this case the fermions occupying negative energy levels
form  1D Fermi liquids, in which   Fermi surfaces  are reduced to a Fermi point
$p_z=0$.   This is very similar to the fermion zero modes in cosmic strings.

Which of three types (a-c)  of the behavior of the fermionic spectrum is
realized depends first on the property of $Q_n$.   In Eq.(6.1.2b)  the quantity
$Q_n$  for the single-particle fermionic excitation  is  integer, if the
winding
number $n$ is even, or half-of-odd-integer if $n$ is odd (note that the
eigenvalues of ${\bf N}$ are $N=1$ for particle and $N=-1$ for the hole).

The state with a vortex line can also possess discrete symmetries
\cite{SalVol,MisVol}. In the most symmetric vortices these are space inversion
symmetry ${\bf P}$, and combined ${\bf TU}_2$ symmetry which
corresponds to the overturn of the vortex axis with the simultaneous time
inversion, the circulation being unchanged under this combined operation. One
more important symmetry is related to the structure of
Bogoliubov fermions: this
is the symmetry under the operation ${\bf C}$ of transformation of Bogoliubov
particle into Bogoliubov hole. Transformations of the quasiparticle spectrum
under these operations are [11]:
$$
{\bf C} E(Q_n,p_z)=  - E(-Q_n,-p_z)~~,
$$
$$
{\bf P} E(Q_n,p_z)= E(Q_n,-p_z)~~,
\eqno(6.2.2)
$$
$$
{\bf TU}_2 E(Q_n,p_z)= E(Q_n, p_z)~~,
$$
where $E(Q_n,p_z)$ denotes the whole set of eigenvalues corresponding to given
$Q$ and $p_z$.

If  ${\bf CP}$ symmetry is satisfied  one has
$$  E_a(Q_n,p_z)=  - E_b(-Q_n,p_z)~~,\eqno(6.2.3)$$
which means that for the state $a$ with given  $Q_n$ and $p_z$ one can find
another state $b$ which has opposite energy $E$ for opposite $Q$ and the
same $p_z$. Just this symmetry together with the index theorem give
the Eq. (6.2.1)
for the fermions on the  vortex with $n=1$ in $s$-wave superconductor. Indeed,
the index theorem implies that the number of the low-lying branches of
fermions
in terms of the continuous  momentum $Q_n$ is $n$ for each spin projection.
This
means that in the $n=1$ vortex there is the only branch  for each of two
spin quantum numbers $S_z=\pm 1/2$. Since different spin projections do
not mix (we neglect here the spin-orbit interaction), it follows from
Eq.(6.2.2)
that each branch is an odd function of $Q_n$:
$$
E_\uparrow(Q_n,p_z)=-E_\uparrow(-Q_n,p_z)~~,~~
E_\downarrow(Q_n,p_z)=-E_\downarrow(-Q_n,p_z)~~,
\eqno(6.2.3)
$$
and one obtains the  Eq.(6.2.1), where $Q_n$ now is the discrete variable. The
factor $\omega_0(p_z)$ is known from the semiclassical limit of large $Q_n$.

The Eq.(6.2.1) can be interpreted as the interaction of the  orbital momentum
$\vec Q=Q_n\hat z$ of the localized fermion with the internal ``magnetic''
field
$\vec h(p_z)=\omega_0(p_z)\hat z$, produced by the vortex:
$$
E=\vec Q\cdot\vec h(p_z)~~.
\eqno(6.2.4)
$$
If the quasiparticle orbital momentum is zero, its energy is exactly zero.

This is not the case for the $n=1$ vortex in $s$-wave superfluids
or superconductors, where $Q_n$ can  only be half-of-odd-integer, which results
in the gap. This small gap in the spectrum can be in principle reduced by some
extra perturbations, such as crystal field, spin-orbit interaction,   external
magnetic field, etc.

\subsection{Fermion Zero Modes in semiclassical approach}

The existence of fermionic zero modes does not
depend on the details of the system and is completely defined by topology.
So here we will consider the simplest and best known case of an
axisymmetric singular vortex in superfluid or
superconductor with $s$-wave pairing. The orbital quantum number $Q_n$ is
considered here as a continuous variable and so one can use the
quasiclassical approximation for the fermions localized in the vortex core.
The Bogoliubov Hamiltonian for the fermions with given spin projection is
a $2\times 2$ matrix
$$
{\cal H}=\check \tau_3\vec q\cdot(-i\vec\nabla)/m +
\check \tau_1 Re\Delta(\vec r)-\check
\tau_2 Im\Delta(\vec r)~.
\eqno(6.3.1)
$$
Here $\vec q=\vec p-\hat z(\vec p\cdot\hat z)$ is the quasiparticle momentum
in the transverse plane and $\Delta(\vec r)=e^{in\phi}\vert\Delta(r)\vert$ is
the gap function (order parameter) in the axisymmetric vortex with winding
number $n$.

The quantum numbers, which characterize the fermonic levels in this
approximation, are (i) the magnitude of transverse momentum  of quasiparticle
$q$, which  is related to the longitudinal projection of momentum
$q^2=p_F^2-p_z^2$,
(ii) the radial quantum number $n_r$ and (iii) the continuous impact parameter
$y=\hat z\cdot(\vec r\times\vec q)/q$. It is  related to the angular momentum
$\hbar Q_n$ by $Q_n=qy$. Introducing the coordinate $x=\vec r\cdot\vec q/q$
along $\vec q$, such that $r^2= x^2+y^2$, and assuming that in the important
regions one has $\vert y \vert\ll \vert x\vert$,  one obtains the
dependence of the
gap function in the singly-quantized vortex ($n=1$) on $x$ and $y$:
$$
\Delta(\vec r)\approx \vert\Delta(\vert x\vert )\vert ( sign(x) -i{y\over
{\vert x \vert}})~~,
\eqno(6.3.2)
$$
and the Hamiltonian:
$$
{\cal H}={\cal H}^{(0)}+{\cal H}^{(1)}~~,
$$
$$
{\cal H}^{(0)}(x)=-i \check \tau_3 {q\over
m}\nabla_x +\check \tau_1 \vert\Delta(\vert x\vert )\vert  sign(x) ~~,
{}~~{\cal H}^{(1)}(x,y)=\check \tau_2 y{{\vert\Delta(\vert x\vert )\vert  }
    \over {\vert x \vert}}~~.
\eqno(6.3.3)
$$
The Hamiltonian ${\cal H}^{(0)}(x)$ is ``supersymmetric'' -
there is an operator $\check \tau_2$ which anticommutes with
${\cal H}^{(0)}(x)$, i.e. $\{ {\cal H}^{(0)},\tau_2\} =0$, and
it has an integrable eigenfunction corresponding to zero eigenvalue:
$$
{\cal H}^{(0)}\Psi^{(0)}(x)=0~~,~~\Psi^{(0)}(x) \propto
(\check \tau_0-\check \tau_2)\exp \left [
{-{m\over q}\int_0^{\vert x\vert }dr~\vert\Delta(r)\vert} \right ] ~~.
\eqno(6.3.4)
$$
Here $\check \tau_0$ is the diagonal $2\times 2$ matrix.

Using first order perturbation theory in
${\cal H}^{(1)}$ one obtains the lowest energy levels:
$$
E(n_r=0,Q_n,p_z)\approx <0\vert {\cal H}^{(1)}\vert 0>=-y
<{{\vert\Delta(r)\vert
}\over r}>=-Q_n\omega_0(p_z)~~,
\eqno(6.3.5)
$$
$$
\omega_0(p_z)={{1}\over q(p_z)}
       {{ \int_0^\infty dr\vert\Psi^{(0)}(r)\vert ^2
{\vert\Delta(r)\vert}/r }\over
{\int_0^\infty dr\vert\Psi^{(0)}(r)\vert^2 }} ~~.
\eqno(6.3.6)
$$

This is the anomalous branch of the low-energy localized fermions obtained
in Ref. \cite{Caroli}. If the energy spectrum is considered as a continuous
function of $Q_n$, this anomalous branch crosses zero at $Q_n=0$.

\

\subsection{Spectral Flow in  Vortices: Callan-Harvey mechanism
of anomaly cancellation}\label{Callan-Harvey}

Now let us consider again the  force, which arise when  the
vortex moves with respect to the heat bath. In Sec.5.3 we discussed this for
the special case of the continuous $^3$He-A vortex  where the  macroscopic
Adler-Bell-Jackiw  anomaly equation could be used. Now we consider this
effect using the microscopic description of the spectral flow of fermion
zero modes within the vortex core. We show that the same force arises for any
vortex in any superfluid or superconductor under a special condition.

If the vortex moves with the velocity $\vec v_L$ with respect to the heat
bath, the coordinate $\vec r$ is replaced by the $\vec r-\vec v_Lt$ and the
impact parameter  $y$ which enters the quasiparticle energy in Eq.(6.3.5)
shifts  with time. So the energy becomes
$$
E_{n_r=0}(Q_n,p_z,t)=-(y- {{\epsilon (\vec q)}\over q} t)q\omega_0(p_z)=-(Q_n
-\epsilon (\vec q)t)~\omega_0(p_z)~~.\eqno(6.4.1)
$$
Here $\epsilon (\vec q)=\hat z\cdot(\vec v_L\times \vec q)$ acts on
fermions localized in the core in the same way that an
electric field acts on the fermions localized on a string in
relativistic quantum theory. The only difference is that under this
``electric'' field the  spectral flow occurs in the $Q_n$ direction rather
than along $p_z$. Along this path the fermionic levels cross the zero
energy level at the rate
$$
\partial_t Q_n=\epsilon (\vec q)=\hat z\cdot(\vec v_L\times \vec
q)~~.
\eqno(6.4.2)
$$
This leads to the quasiparticle momentum transfer from the vacuum (from
the levels below zero) along the anomalous branch into the heat bath.
This occurs at the rate
$$\partial_t\vec P=\sum_{\vec p} \vec p~\partial_t Q_n= {1\over 2}N_{\rm
zm}\int_{-p_F}^{p_F} {{dp_z}\over{2\pi}}
\int_0^{2\pi}{{d\phi}\over{2\pi}}~\vec q
\epsilon (\vec q)=\pi n {{p_F^3}\over {3\pi^2}}\hat z\times\vec
v_L,~~
\eqno(6.4.3)
$$
where the factor ${1\over 2}$ compensates the double counting of particles
and holes,
and we used the index theorem that the number of anomalous
branches (fermion zero modes) is related to the winding number: $N_{\rm
zm}=2n$. This gives the force acting on the vortex when it moves
with respect to the heat bath:
$$
\vec F_{spectral~flow}=\kappa n\hbar C_0 \hat z \times(\vec v_n-\vec v_L)
{}~,~~C_0=
{{m_3p_F^3}\over {3\pi^2}}~~.
\eqno(6.4.4)
$$
Here $\vec v_n$ is the velocity of the heat bath, which in equilibrium
coincides with the velocity of the normal component of the liquid consisting
of the thermal excitations.

Here it is implied that all the quasiparticles, created from the negative
levels of the vacuum state, finally become part of the normal component,
i.e. there is nearly reversible transfer of linear momentum from
fermions to the heat bath. This should be valid in the limit of large
scattering rate: $\omega_0\tau\ll 1$, where $\tau$ is the lifetime of the
fermion  on the $Q_n$ level. This condition, which states that the
interlevel distance on the anomalous branch is small compared to the
life time of the level, is the crucial requirement for spectral flow to
exist. In the opposite limit $\omega_0\tau\gg 1$ the spectral flow is
suppressed and the corresponding spectral flow force is exponentially small
\cite{KopninVolovik}. This shows the limitation for exploring the
macroscopic Adler-Bell-Jackiw  anomaly equation (Eq.(5.1.1)) in the
electroweak model and Eq. (5.2.7) in $^3$He-A.

The reactive force $\vec F_{spectral~flow}$ from the heat bath on the
moving vortex is the consequence of the reversible flux of momentum from
the vortex into the region near the axis, i.e. into the core region.
Within the core the linear momentum of the vortex transforms to the
linear momentum of the fermions in the heat bath when the fermionic
levels on anomalous branches cross the chemical potential.

The process of transfer of linear momentum from the superfluid vacuum
to the  normal motion of fermions within the core is the realization of the
Callan-Harvey mechanism for anomaly cancellation \cite{Callan-Harvey}.
In the case of the condensed matter vortices the anomalies - nonconservation
of linear momentum both in the one-dimensional world of the vortex core
fermions and in the three-domensional Bose-condensate outside the vortex
core - compensate each other. This is the same kind of  the Callan-Harvey
effect which has been discussed in Sec.(5.3) for the motion of
continuous textures in $^3$He-A. $^3$He-A is, however, a very special
superfluid, since due to its internal topology it  always contains the gap
nodes in the spectrum. The nodes lead to momentum   nonconservation, if one
considers the superfluid condensate motion alone.  This is the result of the
transfer of momentum to the normal fermionic  system due to the level flow
through the gap nodes. As distinct from  $^3$He-A, where the gap nodes are
always present, the Callan-Harvey effect for vortices occurs in any
Fermi-superfluid: the anomalous fermionic $Q$ branch, which mediates the
momentum exchange,  always appears in the singular or continuous core, due to
nontrivial  topology of the quantized vortex. This type of Callan-Harvey
effect does  not depend on the detailed structure of  the vortex core and even
on the  type of pairing, and is defined by the vortex winding number $n$. Thus
it  is the same for the singular and continuous vortices, if the condition
$\omega_0\tau\ll 1$ is fulfilled.

This force appears to be similar to the Magnus force - the hydrodynamic force
acting  on the vortex with the winding number $n$ moving in
the ideal liquid (see below). However there is a great difference between
two forces. While the Magnus force $\vec  F_{\rm Magnus}$ is also
proportional to the winding number $n$, it corresponds to the hydrodynamic
momentum transfer between different parts of the vacuum:
it describes the momentum exchange between the coherent
motion of the inhomogeneous vacuum  (the moving vortex) and the mass
flow in the vacuum at infinity. This does not depend on the fermionic
background. On the contrary, $\vec  F_{spectral~flow}$ describes the momentum
exchange between the coherent motion of the inhomogeneous vacuum (vortex)
and the fermionic degrees of freedom. It disappears if the level flow is
suppressed. In the Bose superfluids, like superfluid $^4$He, the
$\vec  F_{spectral~flow}$ is completely absent: there are no fermions. It
is also absent in the node-free A-phase, where the anomaly
parameter $C_0=0$.

\subsection{Three reactive forces acting on a moving vortex}

There are 3 different velocities, which are relevant for
vortex motion: the superfluid velocity $\vec v_s$ (or the velocity of the
vacuum)  far from the vortex; the velocity of the heat bath $\vec v_n$; and
the  velocity of the vortex line $\vec v_L$. As a result, in general
there are 3 different  nondissipative forces acting on the vortex, which we
discuss below.

Here we assume that the condition  $\omega_0\tau\ll 1$ for the spectral
flow is fulfilled. It is important that in this  limit  case the
dissipative (drag or frictional) forces can be neglected  \cite{KopninVolovik}.
The three nondissipative forces are as follows:
$$
\vec F_{nd}=\vec F_{\rm Magnus}+\vec F_{\rm Iordanskii}+
\vec F_{\rm spectral ~flow} ~~,
\eqno(6.5.1)
$$
$$
\vec F_{\rm Magnus}= n\vec \kappa\times \rho(\vec v_L-\vec v_s)~~,
\eqno(6.5.2)
$$
$$
\vec F_{\rm Iordanskii}=n\vec \kappa\times {\breve \rho}_n(T)
(\vec v_s-\vec v_n) ~~ ,
\eqno(6.5.3)
$$
$$
\vec F_{\rm spectral ~flow}=n\vec \kappa\times C_0(\vec v_n-\vec v_L) ~~.
\eqno(6.5.4)
$$
Each of the three forces is of topological origin.

(i) According to the Landau picture of the superfluid liquid (see
Sec.2.1), Eqs.(2.1.6-7)), its motion consists of the motion of the superfluid
vacuum (with the total mass density $\rho$ and the superfluid velocity $\vec
v_s$) and the dynamics of the elementary excitations. The Magnus force in Eq.
(6.5.1) acts on the vortex  if it moves with respect to the  superfluid
vacuum. As before $n$ is the vortex winding number, $\vec\kappa$ is the
circulation vector: for the Fermi (Bose) superfluids $\kappa= \pi  \hbar /m$
($\kappa=2\pi  \hbar /m$), $m$ is the bare mass of the fermion (boson).
In Eqs.(6.5.2-3) $\vec v_s$ is the vacuum (superfluid) velocity far from the
vortex where the $1/r$ contribution from the velocity field circulating around
the vortex can be neglected. The Magnus force comes from the flux of linear
momentum from the vortex to infinity. The topological origin of this force
in condensed matter was discussed in Refs.\cite{Ao-Thouless,Gaitan}.

(ii) The Iordanskii force \cite{Iordanskii,Sonin1} results from the elementary
excitations outside the vortex core: the vortex line produces for them the
Aharonov-Bohm  potential. This force can be obtained as a sum of the forces
acting on the individual particles according to the equation
$$
\partial_t\vec p= (\vec\nabla\times\vec v_s)\times\vec p \ ,
$$
where $\vec p$ is the quasiparticle
momentum and the vorticity
$\vec\nabla\times\vec v_s=n\vec \kappa \delta_2(\vec r)$
is concentrated in a thin tube (vortex core). The
Iordanskii force is thus
$$
- \sum_{\vec p} f~ \partial_t\vec p=-n\vec \kappa\times ~\int
{{d^3p} \over {4\pi^3}}~
f[E_{\vec p}+\vec p\cdot (\vec v_s-\vec v_n)]=n\vec \kappa\times
{\breve \rho}_n(T)(\vec v_n-\vec v_s) \ .
$$
Here $f$ is
Fermi or Bose function depending on the type of the elementary excitations,
which is Doppler shifted due to the counterflow $\vec v_n-\vec v_s$; and
${\breve \rho}_n(T)$ is the density of the normal component, which can be an
anisotropic tensor. The Iordanskii force is the only  nondissipative force  in
Eq.(6.5.1), which depends on temperature $T$.

In the next subsection we will discuss the analogy of
the Iordanskii effect with the Aharonov-Bohm effect for a
spinning cosmic string \cite{CausalityViolation}. Note that
the  Iordanskii force, though related to the particles surrounding the
string, is not the drag (frictional) force on the string from  the particles.
As distinct from the drag force, the  Iordanskii force is invariant under the
time reversal and thus is nondissipative.

The  Aharanov-Bohm interaction of particles with  cosmic strings  was
discussed in \cite{alford}. But for conventional (nonspinning) strings it
gives rise only to the frictional drag force, which should be added to the
drag force caused by to the scattering of particles off the core of the string
\cite{everett}. In condensed matter vortices the drag force is small in the
considered limit $\omega_0\tau\ll 1$ and is neglected here.  For the global
relativistic string the sum of the nondissipative Magnus
and Iordanskii type forces was discussed in \cite{DavisShellard}
and it was found that they arise only if the string is
spinning.

(iii) The third term in Eq. (6.5.1) exists only in fermionic systems,
where the fermion zero modes can exist.  It was found recently that the
spectral flow is unaffected by the temperature $T$, since the temperature does
not change the topology  of the spectrum of fermion zero modes, and thus the
parameter $C_0$ equals  its zero temperature value in Eq.(6.4.4)
\cite{Makhlin-Misirpashaev}. In  the real Fermi-superfluids the existence of
$\vec  F_{spectral~flow}$   is defined by the condition $\omega_0\tau\ll 1$.
At low $T$ this condition can be violated and the discrete character of the
spectrum suppresses the level flow, as a result this force disappears
\cite{Kopnin}, while the Magnus force always  survives. On the other hand, if
the anomalous force exists, it nearly compensates the Magnus force,
since in the real superfluids the mass density $\rho$ is
very close to the anomaly parameter
$C_0=m_3  p_F^3/ (3\pi^2)$. That is why the anomaly within the
core plays a very important part in the vortex dynamics.

In $^3$He-B the anomalous (spectral flow) contribution to
the force has been experimentally observed. When the temperature increases the
regime of the  suppressed spectral flow  transfers to the regime where  it
nearly compensates the Magnus force \cite{KopninVolovik}. This behavior
reproduces the measured temperature dependence of the  force acting on the
vortex \cite{Bevan}.

\subsection{Aharonov-Bohm effect and analog of the spinning string}

To clarify the analogy between  the Iordanskii force and Aharonov-Bohm effect,
let us consider the simplest cases of phonons propagating in the velocity
field of the quantized vortex in the Bose superfluid $^4$He and fermions
propagating in the velocity field  of the quantized vortex in the Fermi
superfluid $^3$He-A. According to Unruh
\cite{Unruh2} the dynamics of the phonons in the presence of the velocity
field is the same as the dynamics of photons in the gravity field. For the
velocity field of quantized vortex  the phonons obey the equation of motion of
the scalar wave in the metric
$$
ds^2=(c^2-v_s^2(r))\left ( dt +{n\kappa\over 2\pi( c^2-v_s^2(r))}
d\phi \right )^2 -
dr^2-dz^2 -{c^2\over c^2-v_s^2(r)}r^2d\phi^2~~.
\eqno(6.6.1)
$$
Here
$c$ is the sound velocity, $\vec v_s=n\hat\phi \kappa/(2\pi r)$ is the
superfluid velocity around the vortex, and  we use the cylindrical system of
coordinates with the axis $z$ along the vortex line.

The same metric takes place for the gapless Bogoliubov fermions propagating
in the field of the axisymmetric phase vortex with $\hat{\bf l}=\hat z$ in
the A-phase of superfluid $^3$He \cite{HydrodynamicAction}. Let us consider
how the Lagrangian for the fermions
$$
{\cal L}=-i\partial_t+ \cal H
\eqno(6.6.2)
$$
with $\cal H$ from Eq.(4.2.12) is modified in the presence of the superfluid
velocity $\vec v_s$
of the vortex. This potential flow flow can be gauged out by applying the
gauge transformation in Eq.(4.2.9) to the Lagrangian. In the vicinity of the
nodes this gives:
$$
{\cal L}=\sum_{N=1}^4{\bf{\check \tau}}_N e^\mu_N(p_\mu-eA_\mu)  \hskip2mm,
\eqno(6.6.3)
$$
where the vierbein give the folowing metric tensor
$$
g^{\mu\nu} =-e^\mu_0 e^\nu_0+\sum_{N=1}^3   e^\mu_N e^\nu_N~~,
\eqno(6.6.4)
$$
with the components
$$
g^{ij} =\sum_{N=1}^3   e^j_N e^i_N=c_\parallel^2\hat {\bf l}_i\hat {\bf l}_j+
c_\perp^2(\delta_{ij}-\hat {\bf l}_i\hat {\bf l}_j)-v_{(s)i}
v_{(s)j}~~,~~g^{00}=-1~~ , ~~g^{0i}=v_{(s)i}
\eqno(6.6.5)
$$
In this case the ``velocity of light'' is anisotropic and its transverse
component $c_\perp=\Delta_0/p_F$ enters Eq. (6.6.1) if only the motion in
the transverse plane is considered.

Far from the vortex, where $v_s(r)$ is small and can be neglected, this metric
corresponds to that of the so called spinning cosmic string. The spinning
cosmic string (see the recent references \cite{Rot.String,RotatingString})
is a string which has rotational angular momentum. The metric in
Eq. (6.6.1) corresponds to a string with angular momentum
$J= n\kappa/ (8\pi G)$ per unit length and with zero mass.

The effect peculiar to the spinning string is the gravitational
Aharonov-Bohm topological effect \cite{CausalityViolation}. Though the metric
outside the string is flat, there is a time difference for the particles
propagating around the spinning string in opposite directions. For the
vortex (at large distances from the core) this time delay approaches
$$
2\tau={2n\kappa\over  c^2}~~.
\eqno(6.6.6)
$$
This asymmetry between the particles moving on different sides of the vortex
is just the origin of the Iordanskii force acting on the vortex in the
presence of the net momentum  of the quasiparticles:
$2\int d^3p/(2\pi)^3~ \vec p f(\vec p) $.

\subsection{Angular momentum anomaly}

As distinct from the linear momentum,  the orbital angular momentum
projection $L_3$ and the generalized angular momentum $Q_n$ in the
axisymmetric environment are quantized integer or half of odd integer
quantities and they can serve  as a model for the quantized and integer
baryonic charge in electroweak theory.

The  symmetry $Q_n$ of different inhomogeneous but axisymmetric vacua
in Eqs.(6.1.2) tells us that $Q_n$ is a conserved quantity. Moreover, since
$Q_n$ is the sum of the generator which correspond to the particle number,
angular and spin momenta, the charge $Q_n$ is  integer (or half of odd
integer) quantum number from the very beginning. Thus the correct calculations
of the total charge of the whole system (vacuum plus excitations) should also
give integer or half integer value.

Eq.(6.1.2) does not mean, however, that the $Q_n$-charge of the vacuum is
always zero. In the pure fermionic description the total charge of the vacuum
is
$$
<{\rm vac}\vert  {\bf Q_n}\vert{\rm vac}>=
\sum_{Q_n,p_z,n_r} Q_n \Theta (-E_{ n_r}(p_z,Q_n))
\eqno(6.7.1)
$$
where $E_{ n_r}(p_z,Q_n)$ are the
energy eigenvalues of Eq.(4.2.10) for fermions  in the axisymmetric field of
the order parameter, $\Theta (-E_{ n_r}(p_z,Q_n))$ is the
Heaviside function which restricts the sum so
that only the negative energy states contribute to
the vacuum charge.  The $Q_n$ charge of the vacuum can be nonzero if some
discrete symmetry is broken and
$E_{ n_r}(p_z,Q_n) \neq E_{ n_r}(p_z,-Q_n)$.

An example of nonzero vacuum charge is presented by   ferromagnets, where the
$SO(3)$ spin rotation symmetry is broken to  $SO(2)$. So the spin projection
$S_3$ is a good quantum number (charge). The absence of the time-inversion
symmetry leads to the net charge  $<{\rm vac}\vert  {\bf S}_3
\vert{\rm vac}>\neq 0$ of  the vacuum, which corresponds to the spontaneous
magnetic moment of the ferromagnet.

Let us consider the $Q_n$-charge of the continuous $n=2$ vortex in
Eq.(3.3.1) for which $\hat l _z (\infty )= -1$. Then according to Eq.
(6.1.2c) this charge corresponds to the ``electric'' charge $Q=L_3-(1/2)N$
of the homogeneous vacuum in which the
vector $\hat {\bf l}$ is oriented along $\hat {\bf l}(r=0)=\hat z$.
This is  a consequence of the fact that the $n=2$ vortex
can be continuously  transformed to this homogeneous vacuum state. For
simplicity the function
$\alpha(r) $, which enters the axisymmetric ansatz for the $n=2$ vortex in
Eq. (3.3.1) will be considered as a constant.

First we use the macroscopic (hydrodynamic) description in terms of
the anomaly equation. In this approach, the orbital momentum $L_3$ of
the vacuum state is
the momentum of the mass current $ \vec j$ in Eq.(5.2.10):
$$
<{\rm vac}\vert {\bf L}_3\vert{\rm vac}>=\int d^3r ~ \hat z \cdot
 (\vec r \times \vec j)  ~.
\eqno(6.7.2)
$$
For the axisymmetric texture, Eq. (2.1.10) for the superfluid vorticity is
simplified:
$$
\vec \nabla\times\vec v_s=-{{\kappa}\over {2\pi r}} \hat z \partial_r \cos\eta
\ .
\eqno(6.7.3)
$$
and  can be integrated, to obtain the velocity  distribution in terms of the
$\hat {\bf l}$ texture.  The resulting velocity field without singularity on
the axis (at $r=0$) is:
$$
\vec v_s=\hat \phi {{\kappa}\over
{2\pi r}}(1-\cos\eta) \ ,
\eqno(6.7.4)
$$
which requires $\hat {\bf l}(r=0)=\hat z$.

The integration of the first two (regular) terms in
Eq.(5.2.10),   after integration by part and using  the boundary condition
$\rho(R)=0$ (there are no particles outside the vessel of radius $R$), gives
the contribution to
$<{\rm vac}\vert {\bf L}_3  \vert{\rm vac}>$:
 $$
{1\over 2m_3}\int d^3r~\rho(r)={1\over 2} N={1\over 2}<{\rm vac}\vert {\bf N}
\vert{\rm vac}> ~~.
\eqno(6.7.5)
$$
Here $N$ is the total number of particles.

As a result the charge $Q_n=L_3-(1/2)N$ of the vacuum is given by the last
term in the current which resulted from the anomaly equation (5.2.7):
$$
\ \ \ \ \ \
<{\rm vac}\vert {\bf Q}_n
\vert{\rm vac}>=-{1\over 2}\int d^3r~C_0~(\hat z
\cdot (\vec r \times \hat {\bf l}))~(\hat {\bf l}
 \cdot (\vec \nabla \times \hat {\bf l} \, \, ))
$$
$$
=- \pi L\int_0^R dr~r^2C_0
\sin^2 \alpha \sin\eta~(\partial_r\eta +
{{\sin\eta~\cos\eta}\over r})  ~,
\eqno(6.7.6)
$$
where $L$ is the height of the vessel and $R$ its radius.

Thus the vacuum accumulates nonzero $Q_n$ charge only if the
$\hat {\bf l}$ texture contains helicity,
$(\hat {\bf l} \cdot (\vec \nabla \times \hat {\bf l} \, \, ))\neq 0$,
in complete correspondence with the
baryonic charge accumulated by the helical gauge fields \cite{tvgf}.
The helicity of the
$\hat {\bf l}$ texture is nonzero only for $\sin \alpha\neq 0$:
$$
\hat {\bf l} \cdot (\vec \nabla \times \hat {\bf l} \, \, ) =
\sin \alpha~(\partial_r\eta + {{\sin\eta~\cos\eta}\over r})  ~.
\eqno(6.7.7)
$$
Also, $\sin \alpha$ is nonzero only  if the parities ${\bf P}$ ($\vec
r\rightarrow -\vec r$)  and $PTU_2$ are broken ($T$ is the time inversion
symmetry and  $U_2$ is the rotation by $\pi$ about transverse axis, see
Eq.(6.2.2)).

One also can rewrite this $Q_n$-charge of the vacuum using the axial anomaly
equation. Again one considers the process in which the
$\hat {\bf l}$-texture  is constant,  but the parameter $p_F$
changes  from zero at $t=t_0$ (when $p_F=0$, the axial field $\vec
A=p_F\hat {\bf l}=0$ and the $Q_n$ charge of the vacuum is zero)  to its
equilibrium value at present time $t$:
$$
<{\rm vac}\vert {\bf Q}_n \vert{\rm vac}>=
-{1\over {2\pi^2}}\int d^3r~\int^t_{t_0} dt~
Q_n(\vec r,t)~(\partial_t \vec A \cdot (\vec \nabla \times
\vec A \, \, ))  ~.
\eqno(6.7.8)
$$
The integrand  describes the transfer of the chiral charge from the vacuum to
the chiral fermionic quasiparticles. Each created  left fermion carries (at
the moment of creation) the local  charge $Q_n(\vec r,t)$ within the
semiclassical (macroscopic) description of the fermions.

Actually on the microscopic level this transfer of the charge $Q_n$
should occur in quantum steps at the moment when the fermionic level
crosses zero energy. Thus one obtains an integer value (or half
integer, if the vortex is
singly quantized), which  in the limit of large $Q_n$ transforms to
the macroscopic expression (6.7.6).

Let us consider how this occurs at large $Q_n$. We consider the evolution of
the $Q_n$ charge of the vacuum during the change of the texture from the
initial state with $\alpha=0$ (and $Q_n=0$) to the final state with finite
$\alpha$. During the evolution of the vortex structure, $Q_n$ levels cross
the zero energy  and   this leads to the acumulation of the   charge $Q_n$ in
the vacuum. The rate of the charge $Q_n$ production  can be found from the
following consideration.  In the limit of large $Q_n$ the spectrum
$E(Q_n,p_z)$ crosses zero as a function of $Q_n$ at some $Q_n(p_z)$.  This
function $Q_n(p_z)$ changes in the process of the modification of the vortex.
Since the states with $Q_n>Q_n(p_z)$ have positive (negative)
energy while the states with $Q_n<Q_n(p_z)$ have negative (positive)
energy, the change of $Q_n(p_z)$ induces the flow of the $Q_n$ levels through
zero with the rate  $\partial_t Q_n(p_z)$.
Since at each event the charge $Q_n(p_z)$ is
transferred from the vacuum to the fermionic degrees of freedom, the total
rate of the charge transfer is
$$
\partial_t<{\rm vac}\vert {\bf Q}_n \vert{\rm vac}>=\sum_{ p_z} Q_n(p_z)
\partial_t Q_n(p_z)~~.
\eqno(6.7.9)
$$
Thus if one starts from the most symmetric vortex and continuously transfers
this state into the vortex with broken symmetry, one obtains the following
general result  for the charge $Q_n$ of the vortex:
$$ <{\rm vac}\vert {\bf Q}_n \vert{\rm vac}>  ={1\over 2}\sum_{ p_z}
Q_n^2(p_z)~~.
\eqno(6.7.10)
$$
This is valid for any axisymmetric vortex in any superfluid and
superconductor.

For the continuous $n=2$ vortex \cite{Kopnin2}
$$
Q_n(p_z)=   r(p_z)~\sin \alpha~\sqrt {p_F^2-
p_z^2}~,
\eqno(6.7.11)
$$
where $r(p_z)$ is the radius at which
$$
\cos  \eta(r)  ={ {p_z}\over {p_F}}~.
\eqno(6.7.12)
$$
The energy levels with the lowest $\vert E_{ n_r}(p_z,Q_n) \vert $
correspond to the radial quantum $n_r=0$ and are given by
$$
E(Q_n,p_z)={\Delta_0\over { p_F~ r(p_z)~\cos\alpha } }~
(Q_n-Q_n(p_z)) ~.
\eqno(6.7.13)
$$
Though $Q_n$ is discrete,   the distance between the $Q_n$ levels $ \Delta_0/
(p_F~ r(p_z)~\cos\alpha )$ is very small compared with the gap amplitude
$\Delta_0$, which means that the effective $Q_n$ is large and can be
considered as continuous. Calculating the sum in Eq.(6.7.10) using the
Eq.(6.7.11) one reproduces the macroscopic result in Eq.(6.7.6)
\cite{OrbitalMomentum}.

\section{Conclusions}

In this article, we have discussed various aspects of $^3$He
and pointed out the similarities to the standard electroweak
model of particle physics. These analogies occur on the level
of symmetry groups and also on the interaction of fermions
with the order parameter. We have discussed the topological
defects that exist in $^3$He-A and have been observed in
the laboratory. The $n=2$ vortex in $^3$He is directly
analogous to the $Z$-string in the electroweak model.
The similarity in the interactions of fermions with
the order parameter leads to similar anomalies in the
two systems. In the electroweak model, baryon number
conservation is anomalous while in $^3$He the conservation
of angular momentum - without accounting for the angular
momentum of the quasiparticles - is anomalous. The
presence of the anomaly can also be deduced from the
existence of fermionic zero modes in both systems.

While there are striking similarities between the two systems,
there are some obvious differences too that cannot be ignored.
A crucial difference at the level of symmetries is that
the symmetries in $^3$He are global but those in the standard
model are gauged. The fermionic degrees of freedom too are
very different -
the standard model has three families each of which has 15 fermions
(including the color degree of freedom) while $^3$He only has 4
fermions. In any case, we do not expect the analogy between the two
systems to be exact. But the presence of certain similarities
in itself should be useful to address certain field theoretic
questions arising in one system in the context of the other
system. For example, calculations in the condensed matter literature
show that the change in the anomalously conserved angular momentum
is always an integer \cite{OrbitalMomentum}
while corresponding calculations in the particle physics
literature find a non-integer value \cite{nc,jgtv}.
We feel that the analogy between the two systems could be
exploited to resolve such issues. In addition, advantage could
be taken of the experiments currently being performed on
$^3$He. For example,
according to \cite{KopninVolovik}, the linear momentum anomaly
caused by the spectral flow of fermions in the vortex has recently been
observed in $^3$He-B by measuring the temperature dependence of the force
acting on moving  vortices \cite{Bevan} (see Sec. (6.5)).

\

\noindent {\it Acknowledgements:}

We thank A. Ach\'ucarro, M.B. Hindmarsh, T.W.B. Kibble, N.B. Kopnin, M.
Krusius and N. Turok for illuminating discussions. We also thank
M. Krusius, \"U. Parts and V. Ruutu for presenting their experimental results
\cite{Avilov,DLtoDUtransformation} prior to publication and E. Thuneberg
for his permission to use some of his Figures. This work
was triggered by the activity during the Topological Defects program
at the Isaac Newton Institute and would not have been done had it not been for
this impetus. G.E.V. was
supported through the ROTA co-operation plan of the Finnish Academy and the
Russian Academy of Sciences and by the Russian Foundation
for Fundamental Sciences, Grant Nos. 93-02-02687 and 94-02-03121.
T.V. was supported in the early stages of this work by a Rosenbaum
Fellowship at the Isaac Newton Institute.

\vfill
\eject


\vfill
\eject

\centerline{\bf Figure Captions}

\begin{figure}[h]\caption[VortexCore] {(a) Core of singular vortex or
string: $Z$-string, Abrikosov vortex in superconducor and vortex in
neutral superfluid Fermi liquid like $^3$He-B. $f(r)$
characterizes the order parameter or
Higgs field. ($f(\infty)=1$.) The $Z$-string and the  Abrikosov
vortex  have two
length scales:  $1/m_{Higgs}$ or coherence length is the scale
for the change of the order parameter;
while $1/m_Z$ is the  penetration length of the magnetic
field $B_Z$  of $Z$-field, which corresponds to the  penetration length
$\lambda$ of the magnetic field in superconductor.
For electrically neutral
superfluid Fermi liquid $\lambda=\infty$.
The profile functions $f(r)$ for the
unit winding $Z$-string, Abrikosov vortex and for the
$n=1$ $^3$He vortex grow linearly with $r$ for small $r$.
(b) Core of the vortex in the A-phase $n=-2$ singular vortex. Note that
for this doubly quantized vortex the function $f_1(r)$ is quadratic at the
origin.   }
\label{VortexCore}
\end{figure}

\begin{figure}[h]\caption[Soliton]{One of many possible structures of the
topological $Z_2$ soliton in $^3$He-A: the $\hat {\bf l}$-soliton. Here the
orbital vector  $\hat {\bf l}$ rotates by $\pi$ in the cross section of the
soliton wall (a), while the vector $\hat {\bf d}$ of the spin
part of the order parameter is kept constant (b).
}
\label{Soliton}
\end{figure}

\begin{figure}[h]\caption[DUContinuousVortex]{(a) The configuration of
the $\hat {\bf l}$-field in the isolated continuous vortex with the winding
number $n=2$ in  $^3$He-A (Figure provided by E. Thuneberg).
It corresponds
to the $\pi_2$ charge $\nu_l=1$ for the $\hat {\bf l}$-field.
The  $\hat {\bf d}$-field either has the same configuration
($\nu_d=1$,  dipole locking of $\hat
{\bf d}$ and $\hat {\bf l}$ fields)
or nearly constant ($\nu_d=0$, the fields
$\hat {\bf d}$ and $\hat {\bf l}$ are unlocked due to
different topological charges). In the typical experimental situations
under axial magnetic field, both $\hat {\bf d}$
and $\hat {\bf l}$ are perpendicular to the field at infinity.
This means that
the axial symmetry is broken for such situations. (b) Elementary cell of
the periodic vortex structure in the rotating vessel in the regime of low
magnetic field (Figure provided by E. Thuneberg). The elementary cell
contains $n=4$ circulation along the cell boundary.
(c) Vortex sheet. The $\hat {\bf l}$ soliton can contain kinks -
lines which separate the sections of the
soliton with different sense of rotation of the $\hat {\bf l}$ vector.
Each kink
represents the vortex with $n=1$ winding number
(first number in the circles).
This vortex is however nonsingular
$\nu=0$, since the $\hat {\bf l}$-vector also
has a winding number around the kink
(second number in the circles). Such
vortices can live only within the soliton.
Under rotation they enter the soliton
one by one from the side wall of container and the vortex sheet arises.}
\label{DUContinuousVortex}
\end{figure}

\begin{figure}[h]\caption[PhaseDiagram]{(a) Experimental phase
diagram of vortex state in the vessel
rotating with angular velocity $\Omega$
in applied magnetic field.  The transition between the vortices with
different topological charges is of first order. Due to the high
metastability the transition lines are not
well resolved. (b) The theoretical
phase diagram calculated by E. Thuneberg.
The vortex sheet becomes absolutely
stable at high rotation velocity.}
\label{PhaseDiagram}
\end{figure}

\begin{figure}[h]\caption[SingularVortex]{ The vortex with $n=1$ and
with constant $\hat {\bf l}$-vector at infinity is singular.
It contains the hard core within which the order
parameter escapes form the vacuum manifold of
the  $^3$He-A. However practically the whole
vorticity $\vec\nabla\times \vec v_s$ is concentrated outside the hard
core in the soft core, where $\hat {\bf l}$-vector sweeps the
half of unit sphere. (Figure provided by E. Thuneberg)}
\label{SingularVortex}
\end{figure}

\begin{figure}[h]\caption[PointDefects]{ The transtion between the vortices
with different topological charges can be mediated by point defects which
carry the deficit of the topological charge. The  hedgehog in the $\hat {\bf
d}$ field is the interface between the dipole-locked and dipole-unlocked
vortices. In the simplest model the interface between the $n=2$ continuous
vortex  and two $n=1$ singular vortices is represented by the hedgehog in the
$\hat {\bf l}$ field, which is similar to the t'Hooft-Polyakov magnetic
monopole with physical string.}
\label{PointDefects}
\end{figure}

\begin{figure}[h]\caption[VortexTransition]{Transtion between the
dipole-locked
and dipole-unlocked vortices is mediated by hedgehog in the
$\hat {\bf d}$ field. Hedgehog is created  on the top or the bottom  of the
vessel,  where the specific point singularity -- boojum -- always exists as a
termination of the continuous vortex (see Ref. \protect\cite{Boojum}). The
motion
of the hedgehog along the vortex line transforms one vortex to another.
}
\label{VortexTransition}
\end{figure}

\begin{figure}[h]\caption[ZeroModes]{ Anomalous branches of fermionic
spectrum: (a) fermion zero modes for $d$ and $u$ quarks in $Z$-string. The
massive branches of spectrum correspond to higher values of the generalized
angular momentum $Q_n$.
(b-d) In condensed matter vortices the distance between
the branches with different $Q_n$ is so
small that one can consider this disrete quantum
number $Q_n$ as continous parameter.
In this case one has the fermion zero modes
in terms of $Q_n$. The spectral flow along $Q_n$ is responsible for the
anomaly in the vortex dynamics which is analogous to the axial anomaly in
particle physics. (b) Spectrum of fermions in the  $n=1$ vortex in $s$-wave
superconductor. (c) Spectrum  of fermions in symmetric $n=1$ B-phase vortex
looks similar to that in $Z$-string. (d) Spectrum  of fermions in symmetric
$n=1$  A-phase vortex: the branch with $Q_n=0$ has no dispersion.
}
\label{ZeroModes}
\end{figure}

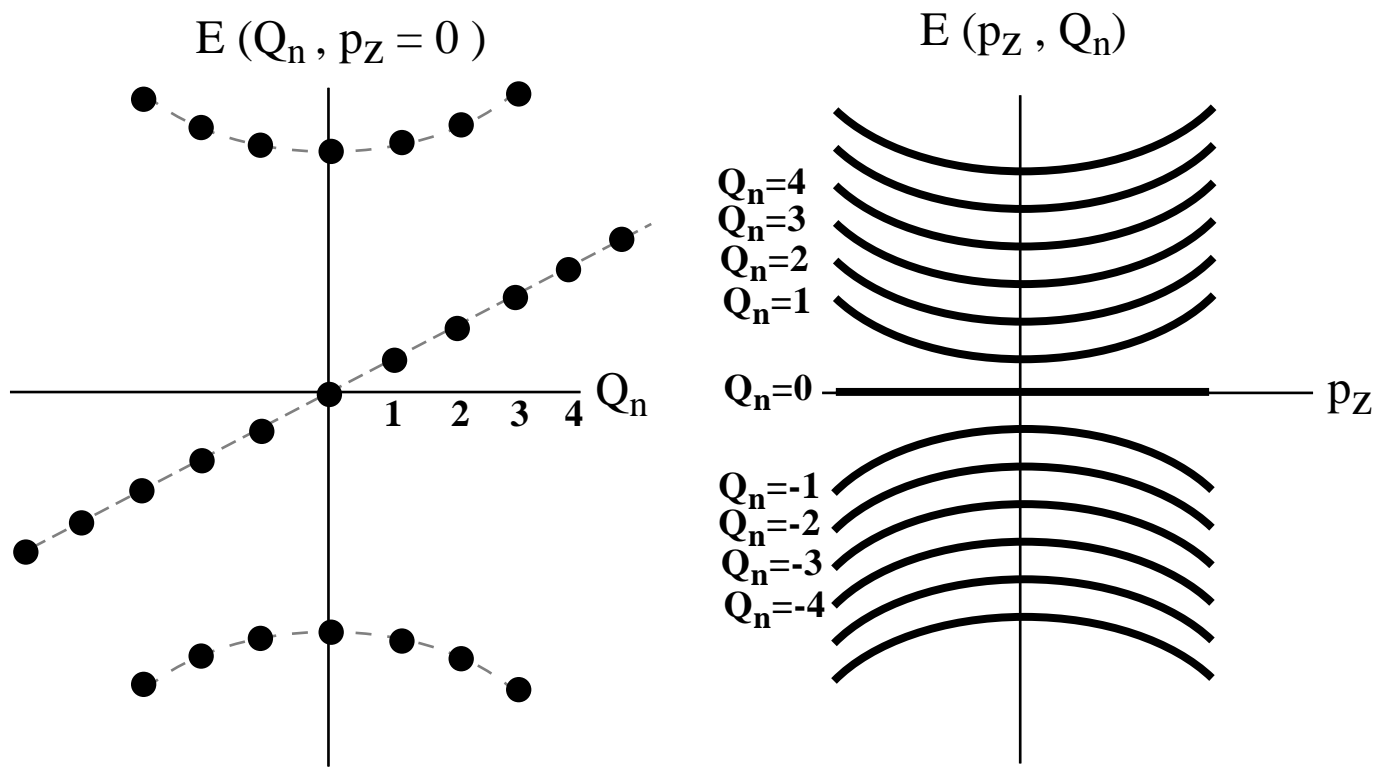
\end{document}